\documentclass[fleqn,usenatbib]{mnras}

\usepackage{newtxtext,newtxmath}
\usepackage[T1]{fontenc}

\DeclareRobustCommand{\VAN}[3]{#2}
\let\VANthebibliography\thebibliography
\def\thebibliography{\DeclareRobustCommand{\VAN}[3]{##3}\VANthebibliography}

\usepackage{graphicx}	% Including figure files
\usepackage{amsmath}	% Advanced maths commands
\usepackage{pdflscape}
\newcommand{\Lxray}{$L_{X}$}
\newcommand{\Lsf}{$L_{SF}$}
\newcommand{\Lbol}{$L_{\rm bol}$}
\newcommand{\EBVstar}{$E_{\rm B - V,\,czt}$}
\newcommand{\EBVagn}{$E_{\rm B - V,\,Pei}$}
\newcommand{\Tgyr}{$\tau_{\rm e}$}
\newcommand{\Agyr}{$t_{\rm age}$}
\newcommand{\Rratio}{$R$}
\newcommand{\angstrom}{\mbox{\normalfont\AA}}
\newcommand{\Laccdisc}{$L_{\rm AGN}$}
\newcommand{\logMstar}{log(M$_*$)}
\newcommand{\NH}{$N_H$}
\newcommand{\opticaldepth}{$\tau_{\rm 9.7}$}
\newcommand{\mcron}{$\mu\rm{m}$}
\newcommand{\ergs}{erg s$^{-1}$}

\defcitealias{Nandra15_AEGISX}{N15}
\defcitealias{Buchner_AEGISX}{B15}
\defcitealias{Ichikawa19}{I19}
\defcitealias{MateoslocalCF}{M16}
\title[Constraining AGN tori at $z\sim 2$ with \textit{JWST}]{Constraining the AGN tori at cosmic noon using high-resolution \textit{JWST} imaging and simultaneous SED fitting}

\author[D. H. Liya et al.]{
Devang H. Liya,$^{1}$\thanks{E-mail: d.h.liya2@ncl.ac.uk}
David J. Rosario,$^{1}$
Matthaios Charidis$^{1}$\\
$^{1}$School of Mathematics, Statistics and Physics, Newcastle University, Newcastle upon Tyne, NE1 7RU, UK
}

\date{Accepted XXX. Received YYY; in original form ZZZ}

\pubyear{2024}

\begin{document}

\label{firstpage}
\pagerange{\pageref{firstpage}--\pageref{lastpage}}
\maketitle

\begin{abstract}
There is evidence for significant evolution in the gaseous and dust properties of galaxies since the era of cosmic noon ($1\lesssim z\lesssim 2.5$). The well known co-evolution of supermassive black holes with their host galaxies suggests a constant connection between the small-scale (nuclear) and large-scale regions of galaxies. A fundamental component of Active Galactic Nuclei (AGN) is the ``torus'', a dense, dusty structure that acts as the interface between the accretion disc and the ISM of the host galaxy. The transitional nature of the torus makes it a prime subject to search for evolution since cosmic noon. We use high-resolution near- and mid-IR imaging from the \textit{JWST} CEERS program to disentangle the emission from the torus in unprecedented detail for 88 X-ray selected AGN at $z\sim2$. We employ a novel SED fitting technique that combines archival low-resolution multi-band photometry at UV to FIR wavelengths with the new high-resolution \textit{JWST} photometry to constrain essential AGN and torus parameters, such as accretion disc luminosity, torus opening angle, and inclination angle. We demonstrate that this SED fitting approach leads to better AGN characterisation and tighter constraints on AGN parameters. The population-level analysis finds that the Covering Fraction ($CF$) distribution peaks at $\approx0.25$ with a long tail towards higher $CF$. Despite the well-known evolution of the ISM and structural properties of AGN hosts to these redshifts, the $CF$ distribution of our sample does not show any strong statistical difference with that found in local AGN of equivalent luminosity, or with those at intermediate redshifts.

\end{abstract}

\begin{keywords}
galaxies: active -- galaxies: Seyfert -- infrared: galaxies -- methods: statistical -- techniques: photometric -- galaxies: nuclei
\end{keywords}

%%%%%%%%%%%%%%%%%%%%%%%%%%%%%%%%%%%%%%%%%%%%%%%%%%

%%%%%%%%%%%%%%%%% BODY OF PAPER %%%%%%%%%%%%%%%%%%

\section{Introduction}
\label{sec:introduction}
Active Galactic Nuclei (AGN) are growing supermassive black holes (SMBHs) at the centres of most massive galaxies in the universe \citep{KormendyHo}. As one of the most energetic phenomena in the universe, they affect the evolution of galaxies through inflows and outflows of material and energy into their hosts \citep[see][and references therein]{AlexanderHickox2012}. Therefore, a thorough understanding of AGN is a crucial part of our picture of galaxy evolution. In the unified picture of AGN \citep{Netzer2015}, there exists an obscuring structure of dense gas and dust -- widely referred to as the ``torus'' because of its predicted shape -- spanning up to tens of parsec \citep{torussize_MIRImage, torussize_MIRInter, torussize_radio} around the central accretion disc. This structure serves as the bridge for material to pass between the wider interstellar medium (ISM) of the host galaxy and the accretion engine close to the SMBH.

The AGN torus in the simplest unification scheme was thought to be an axisymmetric structure around the SMBH, with the observational properties of the AGN only influenced by the luminosity of the central engine and the inclination of the observer's line-of-sight to the axis of the torus \citep{Antonucci1993, UrryPadovani1995}. For the sake of computational simplicity, early models of the torus assumed a structure with uniform density \citep[e.g.][]{smoothAGN}. However, numerous studies have since shown that a better description of a real torus is one with clumps of dusty gas of varying sizes \citep[e.g.][]{clumpyIR, clumpyxray}, possibly embedded in a uniform inter-clump medium \citep[e.g.][]{interclump}, with no clear boundaries marking the edges of the structure. Even though such a structure is not a ``torus'' in the strictly topological sense, we will nonetheless use this term in our work to refer to all these geometries. 

Advancements in simulations over last two decades have led to a number of models that treat the torus as a clumpy structure with various parameters controlling the exact geometry. These models can be divided broadly into static \citep[e.g.][]{Nenkova08, Honig_AGNmodel, skirtor12} or dynamical \citep[e.g.][]{AGNtorus_wada02, AGNmodel_Siebenmorgen, AGNtorus_Wada12} models, respectively based on whether they model the torus as a fixed structure to explain immediate observations, or as a dynamic structure that incorporates the time-dependent physics controlling the formation and evolution of the torus. The first class of models are useful for studying the instantaneous emission from the torus over a broad range of geometries.

A number of studies of different populations of AGN in the local universe, using various flavours of torus models, have revealed that, in contrast to the original canonical torus, the observed properties of AGN also vary because of differences in the geometry of their tori, in addition to differences in inclination and AGN luminosity. Covering fraction ($CF$) -- defined as the proportion of the sky covered by the obscuring torus as seen from the central SMBH -- has emerged as the most important of these geometrical parameters as it shows considerable variety when constrained from the observed emission of AGN \citep[e.g.][]{2011MNRAS.412..835Y, cf_scatter}. A number of studies have identified intrinsic population-level differences between Type-1 and Type-2 AGN beyond just the orientation, the most striking of which is the dichotomy in $CF$ of Type-1 and Type-2 AGN \citep[e.g.][]{RA11, Ichikawa15, MateoslocalCF, GB19}. 

Physically, the torus is thought to be produced and sustained by infalling material from the host galaxy, with a substantial enhancement of dust originating from in-situ formation due to stellar processes in circumnuclear star-forming regions \citep[e.g.][]{schartmann10, AlexanderHickox2012}. It stands to reason that the properties of the dusty torus may be ultimately correlated with the properties of the host galaxy such as the gas fraction and metallicity. Studies have shown that these host galaxy properties at the time of cosmic noon ($z\sim 2$) were significantly different from the local universe \citep{MadauDickinson}, including among AGN hosts \citep[e.g.,][]{Rosario2012, Rosario2015}. Therefore, the properties of AGN tori, especially gross descriptors like covering fraction, might also have evolved between cosmic noon and the local universe. 

A number of X-ray studies, using the fraction of obscured to total AGN as a proxy for $CF$ in a given redshift bin, indeed suggest a higher obscured AGN fraction beyond $z \sim 2$ as compared to the local universe \citep[see][]{merloni14, peca23, Guetzoyan25}. This result is in contrast with Spectral Energy Distribution (SED) studies which define $CF$ for individual AGN based on the ratio of IR to optical emission\footnote{Note that $CF$ obtained from this ratio is normally calibrated to a particular model, and as such is model-dependent.}. While this method has been used extensively in local populations, only a handful of studies such as \cite{toba21} and \cite{trefoloni25} have applied this method to explore the redshift evolution of $CF$. Both these studies find no significant signs of such evolution, suggesting differences in the nature of X-ray vs. IR obscuration.

Under the unification scheme, the dust in the torus absorbs the UV-optical photons from the accretion disc and reprocesses this energy into radiation at infrared wavelengths, particularly in the near-infrared (NIR) and mid-infrared (MIR) in range $1$--$50$ \mcron. This makes NIR and MIR wavelengths the primary probes for the geometry and composition of the dust in the torus. However, using data with $>1$ arcsecond resolutions, the IR emission from the torus is often heavily diluted by dust and stellar emission from the host galaxy, particularly in the more common lower-luminosity AGN and at higher redshifts. This makes it extremely difficult to study the torus emission in isolation for a representative  population of AGN, except in very nearby systems, where the torus emission can be reasonably isolated from the host galaxy using arcsecond-resolution images \citep[e.g.][]{Asmus2014, resolvedtorus_GB21}; or the brightest AGN population (Quasars), where the AGN emission is significantly brighter than the host emission. 

This has been remedied by the advent of the James Webb Space Telescope (\textit{JWST}), which has ushered in an order of magnitude improvement in the resolution and sensitivity of imaging over the Infrared Array Camera (IRAC) and Multiband Imaging Photometer (MIPS) from the \textit{Spitzer} Space Telescope, the erstwhile best resource for NIR/MIR imaging from space. This improvements allow us, for the first time, to separate out the emission from the central kiloparsec of galaxies and AGN around the time of cosmic noon. Large area extragalactic surveys using \textit{JWST} make it possible to separate out and study torus properties in a population of equally-luminous AGN in a significantly different regime from those in the local Universe. In this first exploratory work, we make use of the Cosmic Evolution Early Release Science (CEERS) \citep{CEERS_2017} public data releases to explore the torus properties of a sample of 88 distant X-ray selected AGN.

The IR emission from the dusty torus arises from very small scales ($\lesssim$10 pc) compared to the size of the galaxy (few 1000 pc), or indeed the resolution of IR telescopes. Even with \textit{JWST}, such a structure cannot be resolved except in the nearest of local AGN. A common method for dealing with this limitation is to model various radiative components originating from galactic as well as AGN processes, and disentangle them using Spectral Energy Distribution (SED) fitting. However, most torus models have a large number of parameters with complex degeneracies between them \citep[see][]{Nenkova08, cat3d, skirtor12, cat3dwind}\footnote{Also see the interactive app based on SKIRTOR to visualise these degeneracies: https://skirtor.streamlit.app}. Without the use of appropriate estimation methods, we may derive incorrect estimates of the torus parameters, and, more often, significantly underestimate the uncertainties in these parameters. In this work, we have developed a Bayesian fitting framework to tackle these statistical issues and provide a robust estimate of various parameters, including $CF$, for the AGN population around cosmic noon. We also present a novel fitting technique that utilises both the integrated as well as the nuclear SEDs to reliably disentangle the galaxy and torus components in a sizeable AGN population.

This article is organised in the following manner. Section \ref{sec:observations} describes the different observational datasets and catalogues used in this work along with the sample selection techniques. Section \ref{sec:methods} details our methods to compile and perform photometric measurements, SED models used in this work, and the SED fitting techniques we have employed. We present and discuss the main results in Section \ref{sec:results}, including distributions of various properties across the AGN sample. Finally, we conclude with a summary of findings in this work along with next steps to expand the analysis in Section \ref{sec:conclusions}. Throughout this work, we have assumed a flat $\Lambda_{\rm CDM}$ cosmology with: $H_0 = 69.32$~km Mpc$^{-1}$ s$^{-1}$ and $\Omega_m = 0.2865$. This was done through the ``WMAP9 cosmology'' \citep{wmap9} setting implemented in \texttt{astropy}\footnote{https://docs.astropy.org/en/stable/cosmology/realizations.html\#astropy-cosmology-realizations-wmap9}.

\section{Data sets}
\label{sec:observations}
AGN tori are expected to evolve on short dynamical times ($\sim 1 Myr$) \citep{schartmann10, AGNtorus_Wada12}, much less than the evolutionary timescales explored from galaxy populations sampled at different redshifts. Therefore, to statistically constrain the evolution of tori, we need to study a representative sample of AGN at different redshift intervals. Another challenge is disentangling the emission from various AGN components, produced in the very nuclei of galaxies, from galaxy-wide processes like star formation. This can be mitigated by using multi-wavelength observations to constrain individual components, as different physical processes dominate at different wavelengths.

Keeping these constraints in mind, a sample to study torus evolution must come from a field that samples a large parent population of distant galaxies (several 1000s), but also has archival multi-wavelength observations along with \textit{JWST} imaging. While there are a number of fields that satisfy the first two conditions, the last condition narrows our choices to two fields, namely the Extended Groth Strip (EGS) \citep{EGSoriginalsurvey} and Cosmic Evolution Survey (COSMOS) \citep{COSMOSoriginalsurvey}. The \textit{JWST} imaging in EGS, obtained in Cycle 1 of its operation, is the most mature in terms of data readiness and quality testing after extensive validation from the extragalactic community. Therefore, we have undertaken this work in the EGS field ($\approx$ 100 arcmin$^2$), with the eventual goal of expanding our methodology to the COSMOS field ($\approx$ 0.5 deg$^2$).

This section describes the catalogues and images we have used to select the sample of AGN used in this work, and construct their panchromatic SEDs. The details of the photometric sources used in this work, including information about instruments, filters, key catalogues and references, are given in Table \ref{tab:filters_surveys}.

\begin{table*}
	\centering
	\caption{Summary of all the instruments and filters used to obtain photometric measurements in this work. These observations are described in sections \ref{subsec:CEERS_data} and \ref{subsec:multiband_catalogues}. The construction of SEDs from these catalogues is described in Section \ref{subsec:photometry_method}. While the table gives full set of filters, not every object has an observation in each of these filters, the numbers in the brackets are the number of objects (out of 88) with valid photometric data in that filter. The photometric coverage of the individual sources can be obtained from the accompanying photometric catalogue.}
	\label{tab:filters_surveys}
	\begin{tabular}{llp{0.2\textwidth}p{0.25\textwidth}p{0.25\textwidth}} % four columns, alignment for each
		\hline
        \hline
		Observatory & Instrument & Filters (\# sources) & Primary reference & Photometric catalogue\\
		\hline
        \hline
        \textit{JWST} & NIRCam & F115W (71), F150W (66), F200W (69), F277W (70), F356W (79), F410M (70), F444W (70) & CEERS \citep{CEERS_2017} & This work (Section \ref{subsub:centralSED})\\
        \textit{JWST} & MIRI & F560W (7), F770W (13), F1000W (11), F1280W (13), F1500W (16), F1800W (11), F2100W (5) & & \\
        \hline
        CFHT & MegaCam & u* (52), g' (73), r' (85), i' (88), z' (88) & CFHTLS \citep{Gwyn_CFHTLS} & CANDELS multi-band \citep{Stefanon17_CANDELScat} \\
        CFHT & WIRCam & J (88), H (88), Ks (88) & WIRDS \citep{Bielby_WIRDS} & \\
        \textit{HST} & ACS & F606W (84), F814W (88) & CANDELS \citep{CANDELS2} & \\
        \textit{HST} & WFC3 & F125W (86), F160W (88) &  & \\
        \textit{HST} & WFC3 & F140W (64) & 3D-HST \citep{ThreeDHST} & \\
        \textit{Spitzer} & IRAC & 3.6\;$\mu$m (88), 4.5\;$\mu$m (88), 5.8\;$\mu$m (88), 8\;$\mu$m (88) & \cite{IRAC_SCANDELS, IRAC2} & \\
        \hline
        \textit{Spitzer} & MIPS & 24\;$\mu$m (84) & FIDEL \citep{Fidel} & CANDELS/SHARDS mulit-wavelength catalogue \citep{barro19_CANDELScat} \\
        \hline
        \textit{Herschel} & PACS & 100\;$\mu$m (34), 160\;$\mu$m (42) & - & HELP \citep{HELP_cat}\\
        \textit{Herschel} & SPIRE & 250\;$\mu$m (75), 350\;$\mu$m (75), 500\;$\mu$m (72) & - & \\
        \hline
        \hline 
	\end{tabular}
\end{table*}

\subsection{Sub-arcsecond resolution IR imaging from \textit{JWST}}
\label{subsec:CEERS_data}
The sub-arcsecond \textit{JWST} IR images used in this work were obtained from the CEERS program. This program is a $\approx$100 arcmin$^2$ non-contiguous imaging and spectroscopic survey in the EGS extragalactic deep field, using the Near-Infrared Camera (NIRCam), the Mid-Infrared Instrument (MIRI) imager, and the Near-Infrared Spectrograph (NIRSpec) on-board \textit{JWST}. The observations for this program were carried out over two epochs (June 2022, and December 2022), and with multiple instruments observing in parallel at each telescope pointing \citep[as defined in ][]{CEERS_2017}. Various data products from CEERS have been released publicly including NIR and MIR images taken using NIRCam and MIRI, the products we use in this work.

The NIRCam images were taken over 10 pointings using 6 broadband filters (F115W, F150W, F200W, F277W, F356W, and F444W) and 1 medium band filter (F410M) spanning the wavelength range of about $1.5$ -- $4.5$ \mcron. They were processed using a modified \textit{JWST} Calibration Pipeline described in \cite{Bagley_CEERSNIRCam}, and the final mosaics were produced with pixel scale of $0\farcs03$/pixel and 5$\sigma$ point source depth between 28.5-29.2 AB mag across all filters. We downloaded these mosaics as separate files for each pointing and each filter, from the data releases dr0.5\footnote{https://ceers.github.io/dr05.html} consisting of pointings 1, 2, 3, 6 and dr0.6\footnote{https://ceers.github.io/dr06.html} consisting of pointings 4, 5, 7, 8, 9, 10. 

The MIRI images were taken over 8 different pointings using a total of 7 broadband filters (F560W, F770W, F1000W, F1280W, F1500W, F1800W, and F2100W) with wavelengths ranging between $\sim$5.5 \mcron\;and $\sim$21 \mcron. These images were taken in parallel to the primary NIRCam observations. This setup, coupled with the fact that the MIRI imager's field of view is much smaller than that of NIRCam, results in non-overlapping coverage between these two instruments; some of the area observed by NIRCam does not have MIRI coverage and vice-versa. In addition, none of the pointings have coverage from all MIRI filters; coverage varies between two to six filters per pointing. Interested readers can examine full details of the CEERS observing plan and coverage on the survey website\footnote{https://ceers.github.io/obs.html}. The raw MIRI images were processed using a modified \textit{JWST} Calibration Pipeline described in \cite{Yang_CEERSMIRI}. The mosaics were produced with pixel scale of $0\farcs09$/pixel and 5$\sigma$ point source depth between 22.2-26.2 AB mag across all filters. All these mosaics were downloaded from data release dr0.6. 

In addition to the NIRCam and MIRI mosaics, the CEERS team has also released Hubble Space Telescope (\textit{HST}) optical and NIR images, astrometrically-aligned to the NIRCam images with the same pixel scale. We used these \textit{HST} images primarily to validate our \textit{JWST} photometric measurements, as described later in Section \ref{subsub:centralSED}. In our photometric analysis, we used both the flux mosaics and the error mosaics from the CEERS images (both \textit{JWST} and \textit{HST}), calculated using the reduction pipeline described in the respective data papers.

\subsection{UV-FIR multi-band catalogues}
\label{subsec:multiband_catalogues}
While the \textit{JWST} imaging forms the novel part of our analysis, we build upon the wealth of archival multi-wavelength observations in the EGS field to disentangle various radiative components from the galaxy and the near-nuclear region. As described below, we used three multi-wavelength catalogues produced for this field to construct the SEDs of our sources, spanning UV to FIR wavelengths. 

CANDELS \citep{CANDELS1, CANDELS2} is an \textit{HST} multi-cycle treasury programme that covered a part of the EGS with NIR imaging using the Wide-Field Camera 3 (WFC3). \cite{Stefanon17_CANDELScat} published a multi-band photometric catalogue based on sources extracted in the WFC3/F160W band. The catalogue provides photometric measurements from Canada-France-Hawaii Telescope MegaCam (CFHT/MegaCam), CFHT Wide-field InfraRed Camera (WIRCam), and \textit{Spitzer} InfraRed Array Camera (IRAC) in additon to \textit{HST} Advanced Camera for Surveys (ACS) and \textit{HST}/WFC3. The photometric measurements on all \textit{HST} bands were performed on PSF-matched images to WFC3/F160W band. Sources extracted from WFC3/F160W images were also used as priors to perform photometry on the lower resolution images from \textit{Spitzer}/IRAC and ground-based facilities using \texttt{TFIT} \citep{TFIT}. In the mid-infrared, we have used 24$\mu$m photometry from the \textit{Spitzer} Multi-Band Imaging Photometer (MIPS) imaging in the EGS, compiled in \cite{barro19_CANDELScat}. These measurements were made by direct detection without priors and matched to the WFC3/F160W counterparts as described in the Appendix D of \cite{barro19_CANDELScat}. Both these catalogues were obtained from the Mikulski Archive for Space Telescopes (MAST) portal \footnote{https://archive.stsci.edu/hlsp/candels/egs-catalogs} and combined using the overarching CANDELS \texttt{ID}.

FIR photometry in the EGS is available from the Herschel Extragalactic Legacy Project (HELP) \citep{HELP_cat}. The associated HELP photometric catalogue lists measurements from images taken by the \textit{Herschel} Space Observatory with the Photodetector Array Camera and Spectrometer (PACS) and Spectral and Photometric Imaging Receiver (SPIRE) instruments. These images have significantly lower resolution compared the optical and NIR data in CANDELS (PSF FWHM of \textit{Herschel}/PACS 100 $\mu m$ is $\approx$10'', while PSF FWHM of \textit{HST}/WFC3 F160W is $\approx$0\farcs18). Therefore, the accuracy of the FIR photometry for individual sources is limited by confusion due to severe blending of nearby sources. HELP deals with this issue using the \texttt{XID+} approach, detailed in \cite{XIDplus}\footnote{For clarity, we note that the ``XID'' in \texttt{XID+} is completely unrelated to the term ``XID'' used to identify individual sources elsewhere in this paper.}. Briefly, \texttt{XID+} uses prior information about the positions of sources in a given area from higher resolution images to estimate \textit{Herschel}-based FIR fluxes using a Bayesian framework. The catalogue containing PACS and SPIRE fluxes was obtained directly from the HELP webpage\footnote{https://hedam.lam.fr/HELP/dataproducts/dmu32/dmu32\_EGS/data/\newline EGS.fits\_20180501.fits}, and the individual filters used in our work are listed in Table~\ref{tab:filters_surveys}. In addition to the accurate photometry of FIR-bright sources, the advantage of \texttt{XID+} is that it is able to correctly flag cases where fluxes are unreliable or have large errors, which is commonly the case for sources close the confusion limit (as are most sources in our study). To account for this, we treat FIR photometry in a different way by allowing greater freedom during SED fitting, as discussed in section \ref{subsec:sed_fitting_method}.

We note that the sky area covered by CANDELS and HELP completely encloses the CEERS footprint. This means that all the sources in CEERS can be found in CANDELS and HELP.

\subsection{Chandra X-ray source catalogue}
\label{subsec:xray_obs}
X-ray observations are a widely-used and relatively complete method of selecting both obscured and unobscured AGN at high redshifts \citep{brandt_alexander_2015}. Therefore, we define our sample using AEGIS-X, a Chandra X-ray Observatory/ACIS-I survey of the EGS field taken between December 2007 and June 2009 \citep[N15 hereafter]{Nandra15_AEGISX}. AEGIS-X provides contiguous coverage over $\approx 1000$ arcmin$^2$, completely enclosing the 100 arcmin$^2$ covered by CEERS, with a nominal final exposure of 800ks. 

\citetalias{Nandra15_AEGISX} combined the aforementioned X-ray observations and performed source detection with CIAO to produce a point source catalogue\footnote{https://www.mpe.mpg.de/XraySurveys/AEGIS-X/} with 937 X-ray sources. The catalogue provides X-ray fluxes measured in soft ($0.5-2$ keV), hard ($2-7$ keV), ultra hard ($4-7$ keV), and full ($0.5-7$ keV) bands. It also provides optical/IR counterparts determined using a likelihood ratio method \citep{likelihood_ratio} matched to the Rainbow multi-wavelength catalogue \citep{rainbow_barro11} in EGS. 

\citetalias{Nandra15_AEGISX} also released a redshift catalogue\footnote{https://www.mpe.mpg.de/XraySurveys/AEGIS-X/AEGIS-X\_photoz.html} where each source has an associated spectroscopic or photometric redshift. About a third of these redshifts are spectroscopic, the majority of which were obtained from the Keck DEEP2/3 surveys \citep{keckdeep3, keckdeep4}. The photometric redshifts were estimated using a SED fitting method described in \citetalias{Nandra15_AEGISX} and \cite{salvato11_photz}. Briefly, this method takes into account the optical morphology, optical variability and X-ray flux in order to determine whether to fit a galaxy-dominated \citep{photz_galaxytemplate} or an AGN-dominated \citep{photz_agntemplate} template to the SED from the AEGIS-X multi-wavelength catalogue. These photometric redshifts were released along with their full posterior distribution, though we only use median values in our fits.

Finally, \cite{Buchner_AEGISX} (B15 hereafter) have conservatively classified 553 of the original 937 AEGIS-X sources as AGN using a rest-frame hard-band luminosity cutoff of $> 10^{43}$ \ergs, and released a value-added catalogue\footnote{https://www.mpe.mpg.de/XraySurveys/AEGIS-X/ADD-ONs/Xray\_fit/catalogue\_AEGIS-X.csv} with additional source properties determined by X-ray spectral fitting. The \citetalias{Buchner_AEGISX} catalogue provides improved photometric redshifts when spectroscopic redshifts were unavailable. Other properties in the \citetalias{Buchner_AEGISX} catalogue include estimates of the line-of-sight hydrogen column density (\NH), a measure of X-ray obscuration, and the absorption-corrected rest-frame 2-10 keV X-ray luminosity (\Lxray). We make use of the original source catalogue from \citetalias{Nandra15_AEGISX} as well as the revised AGN catalogue from \citetalias{Buchner_AEGISX} to determine the final properties of our sample. 

\subsection{Sample selection}
\label{subsec:sample}
Starting with 937 X-ray sources from the AEGIS-X catalogue, we applied four filtering steps to construct the final sample used in this work: filtering based on \textit{JWST} coverage, redshift accuracy, the $2-10$ keV X-ray luminosity (\Lxray), and final manual filtering of unusual sources. Each of these steps are described below.

\subsubsection{Filtering based on \textit{JWST} coverage}
The first filtering was done by restricting the sample to only those sources which were observed in CEERS to ensure the presence of high-resolution IR imaging. Since all CEERS mosaics are polygonal and contain the gaps between the detectors, it is possible that a given X-ray source may lie within the bounds of the nominal CEERS mosaic but may not have valid imaging data. To filter for this, we first identified a circular region of radius 2'' in \textit{JWST}/NIRCam and \textit{JWST}/MIRI mosaics centred around the optical/IR counterparts of all 937 X-ray sources. We identified sources that lacked coverage by enforcing a criteria that at least 50\% of all the pixels within the aforementioned 2'' circular region must have valid data (i.e. finite and non-zero flux and error values). This ensures good photometric coverage over the central regions of the selected galaxies. Since CEERS covers only a small portion of the area covered by other surveys (AEGIS-X, CANDELS and HELP), a large number of \citetalias{Nandra15_AEGISX} sources are not observed in CEERS. Consequently, this step resulted in the removal of 794 sources from the parent sample of 937, leaving only 143. For the further analysis of these sources, we extracted $10''\times 10''$ cutouts from the flux and error mosaics, centred at the AEGIS-X counterpart coordinates.

\subsubsection{Filtering based on the accuracy of redshifts}
Large inaccuracies in source redshifts, such as those from unreliable photometric redshifts, can adversely affect SED fitting results by placing the key spectral features at the wrong rest-frame wavelengths. To get around this issue, we applied a second filter where we removed sources with unreliable redshifts. We first compiled our own list of ``best'' redshifts based on different catalogues with the following prioritization: 
\begin{enumerate}
    \item Spectroscopic or photometric redshifts published in \citetalias{Buchner_AEGISX}. This was redshift compilation that we considered of the highest provenance, because the redshifts were cross-validated using X-ray spectral analyses, while taking into account the full posterior distribution from the original photometric redshift estimates. 79/143 sources had redshifts from \citetalias{Buchner_AEGISX}.
    \item Good quality (quality flag $\geq 3$) spectroscopic redshifts from the AEGIS-X redshift catalogue. This was in line with the recommendations made in \citetalias{Nandra15_AEGISX}. All good spectroscopic redshifts in the overlap between \citetalias{Nandra15_AEGISX} and \citetalias{Buchner_AEGISX} are consistent. 28 additional sources not present in \citetalias{Buchner_AEGISX} were assigned spectroscopic redshifts in this way.
    \item Photometric redshifts from the AEGIS-X redshift catalogue in \citetalias{Nandra15_AEGISX}, for remaining 36/143 sources. Since all AEGIS-X sources had photometric redshift estimates, independent of the availability of spectroscopic redshifts, no sources were left without a redshift estimate. 
\end{enumerate}
After compiling these redshifts, we excluded sources with unreliable photometric redshifts by enforcing a criteria that $\Delta z/(1 + z) \leq 0.2$ for any given source, where $z$ is the final adopted redshift and $\Delta z$ is the redshift error. This step removed 31 sources, leading to a reduced sample of 112, spanning a range of redshifts out to $z=3.5$.

\subsubsection{Filtering based on X-ray luminosity}
The third filtering was based on the $2-10$ keV X-ray luminosity (\Lxray) of the source. Specifically, we applied a condition that our sources must have an absorption-corrected \Lxray\;$\geq 10^{42}$ \ergs. This step was taken to remove strong starburst galaxies, in which stellar processes can produce substantial X-ray emission \citep[see][and references therein]{mineo14_XraySF}. Like redshift, in order to determine \Lxray, we made use of both \citetalias{Buchner_AEGISX} and \citetalias{Nandra15_AEGISX} catalogues in the following priority order: 
\begin{enumerate}
    \item \Lxray\;reported in \citetalias{Buchner_AEGISX}. Since this is the unabsorbed luminosity determined by X-ray spectral fitting specifically for each source, this is the most accurate estimate of \Lxray. 
    
    \item Using the reported AEGIS-X flux in the hard X-ray band ($2-7$ keV) flux and the following relationship that invokes a K-correction:
    
    \begin{equation}
        \label{eq:lumincalc}
        L_X ({\rm 2-10\;\rm{keV}}) = \frac{4\pi D^2 F_{\rm e_1-e_2}}{(1+z)^{1-\Gamma}}\times \frac{(10\;\rm{keV})^{1-\Gamma}-(2\;\rm{keV})^{1-\Gamma}}{e_2^{1-\Gamma}-e_1^{1-\Gamma}}
    \end{equation}
    
    \noindent where $D$ is the luminosity distance to a source at a best redshift $z$, $F_{\rm e_1-e_2}$ is flux in the observed-frame energy band $e_1$ keV -- $e_2$ keV, and $\Gamma$ is X-ray spectral index.  The K-correction is based on the assumption that the X-ray spectrum is an unobscured power-law with a fixed spectral index $\Gamma$, and we take a canonical value of $\Gamma=1.7$ for this calculation. The hard X-ray flux is primarily produced by the AGN, and is fairly free of both contamination by hot thermal gas and from X-ray obscuration in high-redshift systems, except in the most heavily obscured cases.
    
    \item Using the full X-ray band ($0.5-7$ keV) from \citetalias{Nandra15_AEGISX} for sources with no hard band detections. We use a modified form of equation \ref{eq:lumincalc} to convert the full-band flux to \Lxray. 
    
\end{enumerate}

After these calculations, we removed one source that had no valid estimate of \Lxray\; and 18 with \Lxray\;$< 10^{42}$ \ergs, resulting in a revised sample of 93 sources.

\subsubsection{Manual filtering} \label{man_filt}
\begin{figure}
    \centering
	\includegraphics[width=\columnwidth]{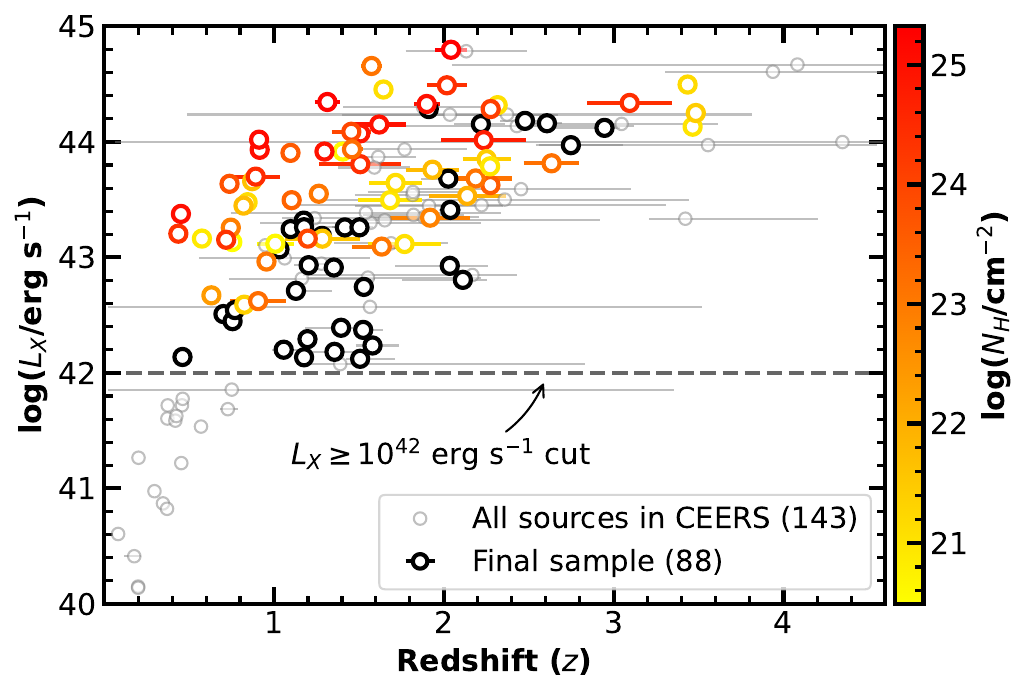}
    \caption{The 2-10 keV X-ray luminosity (\Lxray) plotted against redshift of all 143 AEGIS-X X-ray sources observed in CEERS (smaller gray circles in the background). The 88 sources in our final sample after filtering are shown by larger circles whose colour gives the equivalent Hydrogen column density \NH\; (determined using X-ray spectral fitting by \citetalias{Buchner_AEGISX}) whenever available. These circles are black if the \NH\; is not available. The dashed lines show \Lxray\; cut applied while selecting the final sample (see Section \ref{subsec:sample} for more details).}
    \label{fig:sample_properties}
\end{figure}

Finally, we manually removed 5 more sources due to following reasons. XID408 and XID554 were removed because we could not reliably estimate the errors in their \textit{JWST} photometry from the CEERS images. XID508 was removed because it was deemed to be a star after the examination of its properties in the Rainbow navigator\footnote{https://arcoirix.cab.inta-csic.es/Rainbow\_navigator\_public/} and from its \textit{HST} images.  XID391 and XID506 were removed because their archival photometric measurements were deemed unreliable, after visual inspection of their SED fits (see Section \ref{subsec:example_fits} for more details). This left us with a final sample of 88 sources. Figure \ref{fig:sample_properties} shows \Lxray\; plotted against redshift for our final sample. The points are coloured by X-ray column density \NH\;if available. Table \ref{tab:sample_prop} lists a set of important properties of all these sources. An electronic version of this table with additional columns is available in the supplementary material provided with this paper.

\subsection{Source re-centring in \textit{JWST} images}
\label{subsec:recentring}
In this work, we use small-aperture photometric measurements of the nuclei of our AGN, obtained from the \textit{JWST} images, to isolate the AGN emission. The accuracy of measurements is heavily dependent on how accurately we can determine the location of the AGN nuclei at the centres of their host galaxies. While the optical/IR counterpart coordinates provided in AEGIS-X are largely accurate, there are minor offsets between these coordinates and the centres of the galaxies as seen in CEERS images. This is especially pronounced in irregular galaxies or those with close neighbours. Therefore, we re-centred all sources using \textit{JWST} images to obtain a list of new source coordinates. We call these ``IR centres'' and use them as source coordinates for all the further analysis unless stated otherwise. 

The re-centring procedure was as follows. For each object, we identified the filter with the shortest wavelength that sampled the SED at a rest-wavelength $>2$ \mcron\, which maximises the relative AGN torus emission because the host galaxy stellar emission dips at these wavelengths. In practice, this choice selected a filter with a mean wavelength closest to the observed-frame 5 \mcron: the F444W NIRCam filter in most cases, or the lowest available wavelength MIRI filter in a handful of cases. Once this filter was identified for a given source, we determined the centre using the \texttt{centroid\_sources} function from the \texttt{photutils} (version 1.9.0) Python package\footnote{https://photutils.readthedocs.io/en/stable/index.html} \citep{photutils}. Within \texttt{centroid\_sources}, we used a circular footprint of 1'' and the \texttt{centroid\_2dg} algorithm that fits a 2D Gaussian to the emission in that footprint. These new centres were then visually examined to verify their accuracy and we manually assigned centres for 11 sources where the centres found by the centroiding algorithm were incorrect. Almost all the these have two or more objects within 2''. For 5 of these we assigned the centre coordinate to the best source based on the SEDs and brightness of all the sources in the field instead of the source closest to the CANDELS coordinates. Overall, we found that the median offset between the original AEGIS-X counterpart coordinates and our coordinates was  $\approx 0\farcs2$, with 3 sources having offset between $0\farcs6$ and $1\farcs0$.

\begin{figure*}
    \centering
	\includegraphics[width=\textwidth]{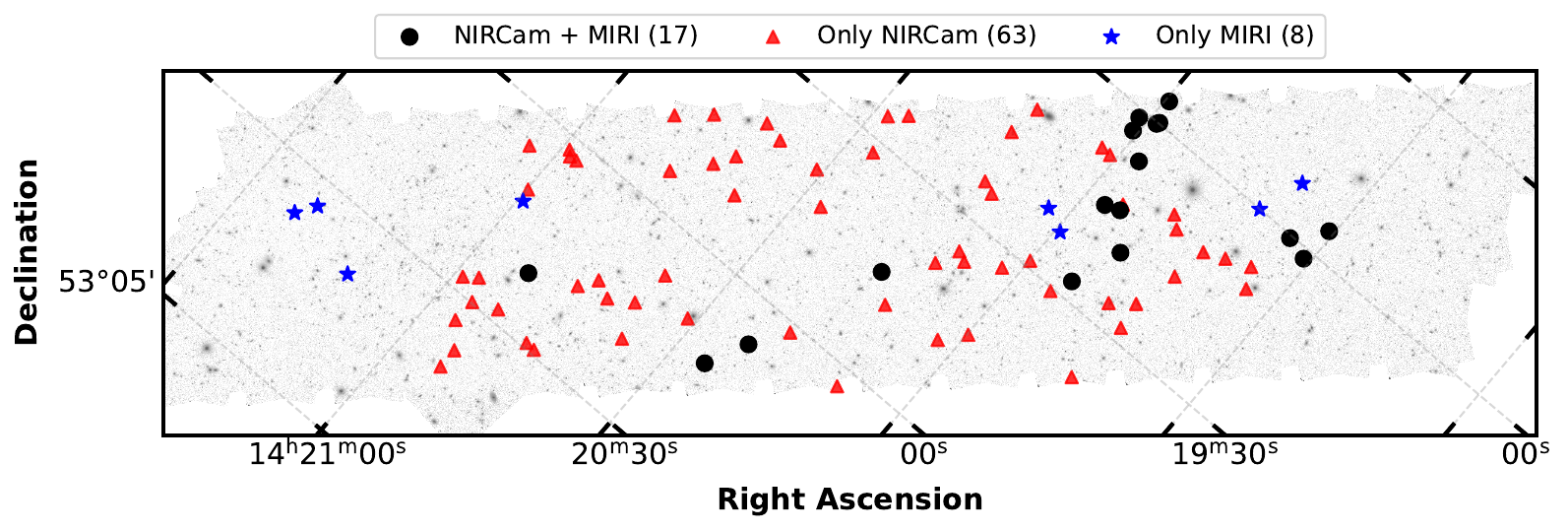}
    \caption{The position of the 88 sources in the final sample used in this study along with the \textit{HST}/WFC3 F160W mosaic from CANDELS in the background. Different symbols show whether a particular source was observed using at least one filter in NIRCam (red triangle), MIRI (blue star), or both (black circle). Note that the region shown in the figure is entirely covered by the HELP and the AEGIS-X surveys.}
    \label{fig:field}
\end{figure*}

The positions of the final sample along with their \textit{JWST} coverage are shown in Figure \ref{fig:field}, relative to the \textit{HST}/WFC3 160W mosaic from CANDELS\footnote{https://archive.stsci.edu/hlsps/candels/egs/egs-tot/v1.0/60mas-all/hlsp\_candels\_hst\_wfc3\_egs-tot-60mas\_f160w\_v1.0\_drz.fits}. 63 sources from the final sample were observed in at least one \textit{JWST}/NIRCam filter, 9 in at least one \textit{JWST}/MIRI filter, and 17 in at least one filter from both \textit{JWST}/NIRCam and \textit{JWST}/MIRI instruments.

\section{Methods}
\label{sec:methods}
\begin{figure}
    \centering
	\includegraphics[width=0.8\columnwidth]{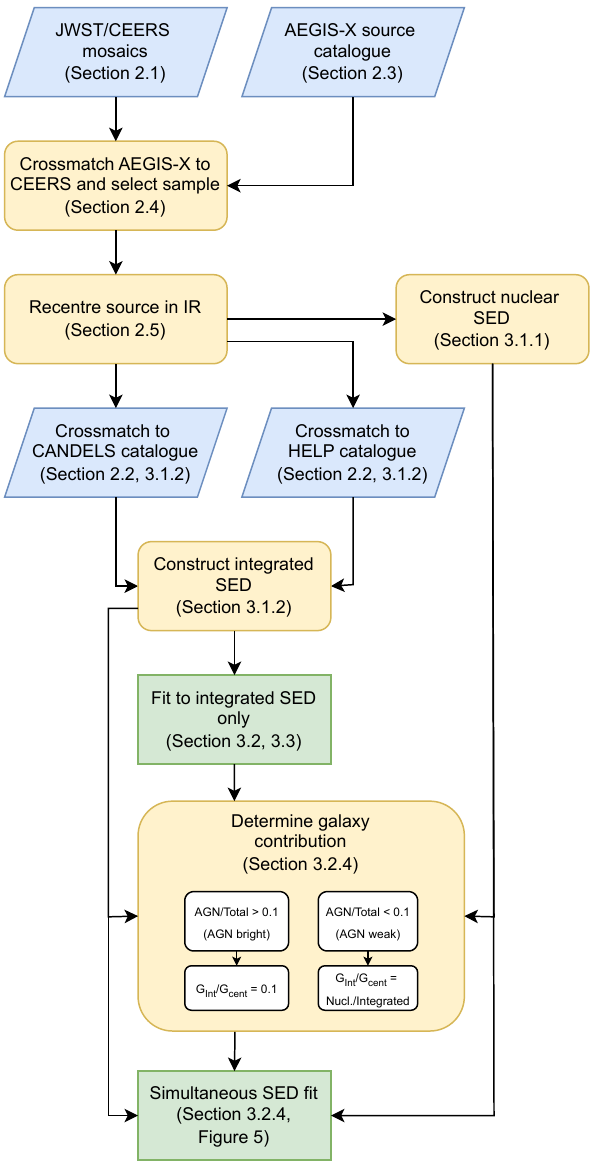}
    \caption{A flowchart describing the workflow of the analysis in this work. The blue boxes represent data product acquired from other studies (e.g. source catalogues, images, etc.), yellow rounded rectangles represent the processing steps, and the green rectangles represent the fitting steps. The arrows indicate the flow of products from one step to others and the section numbers given in the bracket describe those steps in more detail in the main text.}
    \label{fig:flowchart}
\end{figure}

In this work, we construct two sets of SEDs: (i) NIR to MIR SED of only the central region of galaxies measured from high-resolution \textit{JWST} imaging, and (ii) UV to FIR SED, constructed from lower resolution images, in which flux measurements are integrated over the entire galaxy. The construction of these two SEDs is discussed in Sections \ref{subsub:centralSED} and \ref{subsub:integratedSED}, respectively. We then adopt a two-step SED fitting process. In the first step, we fit only the integrated SED to determine the relative galaxy contribution. In the second step, we simultaneously fit both central and integrated SEDs where the AGN component is common to both SEDs, while the galaxy component is scaled for the central SED as informed from the first step. The three model components used in SED fitting are described in Sections \ref{subsub:stellar_pop}, \ref{subsub:AGN}, and \ref{subsub:galaxy_dust}. The SED fitting procedure itself is detailed in Section \ref{subsub:finalSEDmodel}. This approach ensures self-consistent modelling of emission across different physical scales. The Bayesian SED fitting framework along with the priors are discussed in Section \ref{subsec:sed_fitting_method}. Finally, we show some of the examples and discuss a handful of failure cases of our fitting approach in Section \ref{subsec:example_fits}. The Figure \ref{fig:flowchart} summarises the workflow of our analysis with references to the relevant sections in the text.

\subsection{Photometric measurements}
\label{subsec:photometry_method}
The two sets of SEDs used in this work were constructed in the following manner.

\subsubsection{Central IR SED}
\label{subsub:centralSED}
 We measure the central photometry of our AGN hosts with the assumption that it contains emission from an AGN point source that includes the torus and accretion disc. We did this by measuring the flux within a small circular aperture at the IR centres, and then using aperture corrections valid for a point source to estimate the total point source emission. Note that this aperture-corrected flux is still likely to contain a modest amount of light from the inner galaxy that is not related to the AGN. Regardless, for most of our sample, this approach yields a much purer measure of the torus emission than the flux integrated over the entire galaxy. The circular aperture used to measure the central flux had a fixed radius across all filters to ensure that the measured flux comes from a similar physical scale. This radius was $0\farcs 179$, chosen to be comparable to the width of the PSF core at shorter \textit{JWST} wavelengths but large enough to measure the PSF core accurately at longer wavelengths. Specifically, this aperture encloses 80\% of the point source energy in the NIRCam/F277W filter and encloses about 9-11 pixels ($\approx 12$\% of the point source energy) in the longest wavelength MIRI/F2100W filter. 

\begin{figure*}
    \centering
	\includegraphics[width=\textwidth]{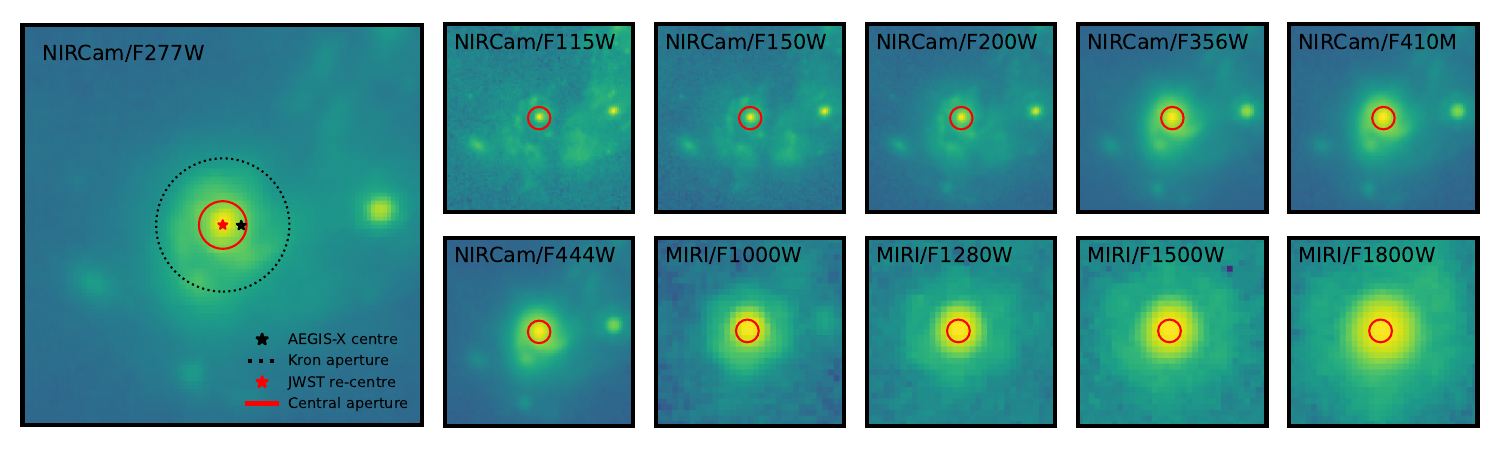}
    \caption{$3''\times 3''$ images of one of our objects (XID471) in different NIRCam and MIRI filters. The large left-most panel shows the image in NIRCam/F277W filter, where the PSF has a 80\% enclosed energy radius of $0\farcs 179$ that was adopted as the aperture size to measure the ``central SED'' (Section \ref{subsub:centralSED}). This aperture was fixed for all filters and is shown by a red circle in each image. The black dotted line in the left-most panel gives the Kron radius of this source used to measure the flux reported in the CANDELS catalogues that form our ``integrated SED'' (Section \ref{subsub:integratedSED}). The black star shows the coordinates of the optical counterpart given by AEGIS-X, which is generally not centred on the galaxy nucleus. We have estimated the centres of all our sources using \textit{JWST} images as described in Section \ref{subsec:recentring}. This new centre is shown by a red star in the left-most panel and used as source co-ordinate throughout this work.}
    \label{fig:example_galaxy}
\end{figure*}

The corrections applied to the aperture fluxes to estimate the total point source emission were derived from model \textit{JWST} PSFs for each NIRCam and MIRI filter, created using WebbPSF\footnote{https://webbpsf.readthedocs.io/en/latest/} \citep{webbpsf}. The fixed aperture used to measure the central flux is shown by solid red lines in Figure \ref{fig:example_galaxy} plotted over multi-band images of an example object from our sample. The correction factors are listed in Table \ref{tab:correction_factors}. This procedure gives an IR SED of the central region of the galaxy which we refer to as the ``central SED'' throughout the paper\footnote{Note that the angular size distance changes very little from $z=0.8$ ($\approx 7.6$ kpc/'') to $z=2.5$ ($\approx 8.2$ kpc/''). So a fixed aperture covers almost the same physical scale across all our targets.}.

The CEERS data release also consists of \textit{HST}/ACS and \textit{HST}/WFC3 images astrometrically aligned to the NIRCam images. Since the resolution of these images is comparable to the NIRCam, we have used the same aperture described above to measure the central flux in optical and NIR bands. While we do not use these fluxes in our SED fitting, they are used to visually validate the continuity of the central SEDs measured from the \textit{JWST} images.

\begin{table}
    \centering
    \caption{Aperture correction applied to the measured flux in a given filter if the flux is measured using a circular aperture of radius 0\farcs179 (80\% of point source energy enclosed in NIRCam/F277W filter).}
    \label{tab:correction_factors}
    \begin{tabular}{lc}
        \hline
        \hline
        Filter name & Correction factor \\
        \hline
        \hline
        NIRCam/F115W & 1.21 \\
        NIRCam/F150W & 1.22 \\
        NIRCam/F200W & 1.22 \\
        NIRCam/F277W & 1.28 \\
        NIRCam/F356W & 1.41 \\
        NIRCam/F410M & 1.44 \\
        NIRCam/F444W & 1.45 \\
        \hline
        MIRI/F560W & 1.81 \\
        MIRI/F770W & 2.00 \\
        MIRI/F1000W & 2.43 \\
        MIRI/F1280W & 3.45 \\
        MIRI/F1500W & 4.5 \\
        MIRI/F1800W & 6.17 \\
        MIRI/F2100W & 8.2 \\
        \hline
        \hline
    \end{tabular}
\end{table}

\subsubsection{Integrated UV-FIR SED}
\label{subsub:integratedSED}
CEERS is completely within the CANDELS survey as well as the AEGIS-X and HELP surveys, which means we can construct a multi-band SED for all sources ranging from the UV to the FIR. We first identified each of our sources in various multi-band source catalogues described in Section \ref{subsec:multiband_catalogues}. This identification was done by finding the closest catalogue source to the re-centred source coordinates described in section \ref{subsec:recentring}. Once the cross-identifications were made, we compiled the photometry from different bands across the three catalogues to construct the final SED. Since these multi-band fluxes were measured for the entire galaxy from low-resolution images, they contain light from the AGN as well as all of the host galaxy. We refer to this set of SEDs as ``integrated SEDs'' throughout this work.

\subsection{SED modelling}
\label{subsec:sed_modelling_method}
In our analysis, we assume that the emission from a given galaxy can be decomposed into constituent SEDs from three different astrophysical components. These components are broadly, (i) Photospheric emission from stellar populations, (ii) AGN emission from the accretion disc and torus, and (iii) Emission from cold dust in the galaxy heated by star-formation. We use this assumption to model our SEDs as a sum of these three components. Within the fitting framework used in this work, each component is defined by a single ``scale parameter'' that sets the normalisation of the model SEDs, and a set of ``shape parameters'' that define the actual shape of the SED. All the free parameters for each model component are summarised in Table \ref{tab:model_params} for quick reference. We discuss each of these components and the associated SED model below.

\begin{table*}
	\centering
	\caption{A summary of the models and corresponding parameters used in this work. Bold parameters are scale parameters while all others are shape parameters (refer to section \ref{subsec:sed_fitting_method} for details). Three different section of this table correspond to components representing (i) Photospheric emission from stellar populations, (ii) AGN emission from accretion disc and torus, and (iii) Emission from cold dust in the galaxy heated by star-formation respectively. The final model used in SED fitting is the sum of these three components. Refer section \ref{subsec:sed_modelling_method} for more information.}
	\label{tab:model_params}
	\begin{tabular}{llclp{0.38\textwidth}p{0.16\textwidth}} % four columns, alignment for each
		\hline
        \hline
		Parameter name & Unit & Range & Prior$^*$ & Short description & Reference(s)\\
		\hline
        \hline
		  $\mathbf{log(M_*)}$ & M$_{\odot}$ & - & U(8.0, 13.0) & Stellar mass of the galaxy & \cite{BC03}, \cite{calzetti00}\\
        $E_{\rm B-V, czt}$ & mag. & 0.0-2.0 & HG(0.0, 0.2) & Extinction coming from the attenuation of star-light by the dust in the galaxy & \\
		\Agyr & Gyr & 0.005-11.0 & U(0.1, cosmo) & Time at which star-formation is assumed to begin & \\
        \Tgyr & Gyr & 0.01-15.0 & U(0.01, 15.0) & SFR decay timescale & \\
		\hline
        $\mathbf{L_{AGN}}$ & \ergs & - & G(X-ray, 0.3) & Luminosity of the unobscured accretion disc at 2500\angstrom & \cite{skirtor12, skirtor16}, \cite{Pei92} \\
        $E_{\rm B-V, Pei}$ & mag. & 0.0-4.0 & HG(0.0, 1.0) & Additional SMC-like extinction of AGN along the line-of-sight &  \\
        $i$ & de.g. & 0.0-90.0 & U(0.0, 80.0) & Inclination angle of the torus i.e. the angle between the line-of-sight and the polar axis &  \\
        $oa$ & de.g. & 10.0-80.0 & U(10.0, 80.0) & Opening angle of the torus as measured from equatorial plane toward the polar axis &  \\
        $R$ & - & 10.0-30.0 & U(10.0, 30.0) & Ratio of outer to inner radius of the torus &  \\
        $p$ & - & 0.0-1.5 & U(0.0, 1.5) & Power-law exponent for the dust density in radial direction &  \\
        $q$ & - & 0.0-1.5 & U(0.0, 1.5) & Power-law exponent for the dust density in polar direction &  \\
        \opticaldepth & - & 3.0-11.0 & U(3.0, 11.0) & Average edge-on optical depth at 9.7$\mu$m &  \\
		\hline
        $\mathbf{L_{SF}}$ & \ergs & - & U(42, 47) & IR luminosity of dust heated by star formation model, integrated between $8-1000\mu m$ & \cite{Dale14} \\
        $\alpha_{\rm SF}$ & - & 0.0625-4.0 & G(2.0, 0.3) & Parameter controlling the relative contributions of dust at various temperatures &  \\
        \hline
        \hline 
        \multicolumn{6}{l}{\footnotesize $^*$ U($a, b$)   : Uniform prior between $a$ and $b$}\\
        \multicolumn{6}{l}{\footnotesize \; G($a, b$)   : Gaussian prior with mean=$a$ and standard deviation=$b$}\\
        \multicolumn{6}{l}{\footnotesize \; HG($a, b$): Half-Gaussian prior with centre at $a$ and standard deviation=$b$}\\
        \multicolumn{6}{l}{\footnotesize \; cosmo: Cosmological age at given redshift}\\
        \multicolumn{6}{l}{\footnotesize \; X-ray: Value based on \Lxray\; as explained in Section \ref{subsec:sed_fitting_method}}
	\end{tabular}
\end{table*}

\subsubsection{Photospheric emission from stellar populations}
\label{subsub:stellar_pop}
Stellar emission from the galaxy is modelled using the stellar population synthesis framework (GALAXEV) from \cite{BC03} which uses observed as well as synthetic stellar spectra. In order to generate our templates, we adopted a Chabrier Initial Mass Function (IMF) \citep{Chabrier03b}; solar metallicity since massive galaxies, such as AGN hosts, have metallicities close to or super-solar even at $z\sim2$; and exponentially-decaying delayed star-formation history \citep[see][and references therein]{Conroy13}. We then apply the Calzetti attenuation law \citep{calzetti00} using the \texttt{extinction} (version 0.4.6) Python package\footnote{https://extinction.readthedocs.io/en/latest/} \citep{extinction_package} with $E(B-V)$ between 0-2 to simulate the attenuation of starlight by the dust in the galaxy, assumed to act as a screen. The free shape parameters of the full dust-attenuated stellar population model are: (i) the time of the beginning of star formation (\Agyr), (ii) the decay timescale of the star formation rate (\Tgyr), and (iii) dust extinction (\EBVstar). The scale parameter for this model is the total stellar mass in the units of $\log$ solar mass (\logMstar). The model SEDs were generated for 10 different values of \Agyr\footnote{5 Myr, 25 Myr, 100 Myr, 290 Myr, 640 Myr, 900 Myr, 1.4 Gyr, 2.5 Gyr, 5 Gyr, 11 Gyr} and 12 different values of \Tgyr\footnote{10 Myr, 50 Myr, 100 Myr, 200 Myr, 500 Myr, 750 Myr, 1 Gyr, 2 Gyr, 5 Gyr, 7 Gyr, 10 Gyr, 15 Gyr}. These model SEDs covered a wavelength range of 91 \angstrom\; to 160 $\mu$m, the GALAXEV default, comfortably larger than the full range of wavelengths where this stellar emission could be dominant.

\subsubsection{AGN accretion disc and torus emission}
\label{subsub:AGN}
We model the UV-FIR AGN emission using SKIRTOR, which is a modern and well-tested radiative transfer-based clumpy torus model \citep{skirtor12, skirtor16}. The AGN torus in SKIRTOR is represented by a two-phase medium consisting of high-density clumps of dust embedded in a continuous low-density medium. It is illuminated by a central accretion disc modelled as a point source with an intrinsic emission spectrum given by a piece-wise combination of 4 power-laws, and an anisotropic illumination pattern proportional to $\cos\theta (2\cos\theta + 1)$, where $\theta$ is the angle from the axis of the accretion disc \citep{Netzer87}. There is some evidence that a two-phase clumpy model of the torus is a better descriptor of AGN than a purely clumpy model \citep[e.g.][]{Nenkova08}. As discussed in \cite{skirtor16}, two-phase clumpy torus models are more consistent with observational evidence from variability studies \citep[e.g.][]{Risaliti2002}, and are better at reproducing the relative NIR luminosity of the AGN, and observed silicate feature strengths.

The SED library, obtained from the SKIRTOR website\footnote{https://sites.google.com/site/skirtorus/sed-library}, consists of 19,200 individual SEDs\footnote{These SEDs are visualised here: https://skirtor.streamlit.app}. We apply additional extinction ($E(B-V)$ in the range 0-4) from a screen of dust to simulate any additional obscuration of the AGN along the line-of-sight from material outside the torus, such as from dust lanes in a host galaxy \citep[e.g.,][]{dustlane}. In line with observational results for QSOs \citep{smcQSO}, an SMC-like law was chosen for the additional extinction.

The free shape parameters for this extinguished SKIRTOR model are: (i) Opening angle of the torus ($oa$), (ii) Inclination of the torus axis from the line-of-sight ($i$), (iii) Average edge-on optical depth of the torus at 9.7 micron ($t$), (iv) ratio of the outer radius to the inner radius of the torus ($R$), (v) the powerlaw exponent for the radial dust density gradient ($p$), (vi) the powerlaw exponent for the polar dust density gradient ($q$), and (vii) additional line-of-sight extinction (parametrised by the reddening \EBVagn). The scale parameter for the model is the intrinsic (unobscured) 2500\angstrom\ accretion disc luminosity (\Laccdisc).

We also define the fraction of sky covered by the AGN torus as seen from the central SMBH to be a new derived parameter ``covering fraction'' ($CF$). Mathematically, this is given by $CF = \cos(90 - oa)$, where $oa$ is the opening angle defined by SKIRTOR and measured from equator towards the pole. This is sometimes referred to as the ``geometrical covering fraction''. Note that the geometrical covering fraction is independent of the inclination angle of the torus.

\subsubsection{Cold dust heated by star-formation}
\label{subsub:galaxy_dust}
The infra-red emission from the cold dust in the galaxy heated by star formation activity is modelled by the two-parameter family of AGN-free templates from \cite{Dale14}\footnote{http://physics.uwyo.edu/~ddale/research/seds/seds.html}. These templates were constructed assuming that galaxies exhibit a power-law distribution of interstellar radiation intensities that heat dust locally, and the emission from the entire galaxy is the combination of these local dust SEDs at different temperatures. The model has one shape parameter, $\alpha_{SF}$, which controls the relative contributions of the local dust SEDs at different temperatures. The scale parameter is the IR luminosity from galaxy dust integrated over the wavelengths of $8$--$1000$ \mcron\ (\Lsf) in units of $\log$ \ergs.

\subsubsection{Combined SED model}
\label{subsub:finalSEDmodel}
We construct the final model as a sum of the three components described above. The two different SEDs we have constructed, integrated and central, will, in general, have different contributions from the galaxy components, but will share the same AGN component. We leverage this fact to devise a two-step fitting process in this work. In the first step, we fit only the integrated SED to determine relative galaxy emission in the central SED. We then use this to set a fixed scaling of galaxy components in our second simultaneous SED fit to the integrated and the central SED. We make the following two assumptions that allows the second simultaneous SED fit.

Firstly, we assume that all of the AGN emission is concentrated within the centre of the galaxy and specifically within the $0\farcs179$ radius aperture used to get the central photometry. This is easily justified because the torus is significantly smaller ($\approx 10$ pc) than the projected aperture size at the redshifts of our sources ($\approx 2$ kpc). This means that the same AGN model component can be fit to both integrated and central SEDs.

Secondly, we assume that within the central aperture, there can be a substantial contribution from stellar light and cold dust heated by star-formation. We account for this emission by scaling these two components by a factor $C \leq 1$ between the integrated SED and the central SED. The exact value of $C$ is determined from the first fit as discussed later in this section.

\begin{figure}
    \centering
	\includegraphics[width=\columnwidth]{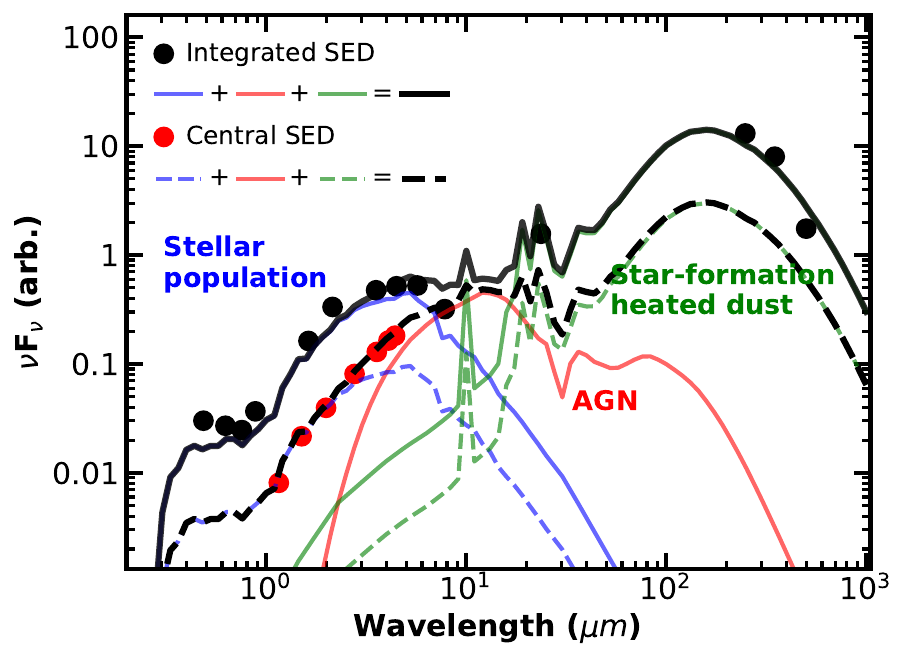}
    \caption{This figure illustrates the concept of simultaneous SED fitting used in this work. Solid and dashed black lines show the total model SEDs for integrated fluxes and central fluxes respectively. These observed fluxes are shown as black and red points without error bars for readability. The stellar population (blue) and dust illuminated by stars (green) models are scaled by a factor of 0.23 (specific to this source) for central SED as compared to the integrated SED, while the AGN model (red) remains same across these two SEDs as indicated by lack of dashed red line. This is mathematically represented by Equation \ref{eq:SEDfit}. Note that for the sake of decluttering we drop all the dashed lines from other similar SED fitting figures in this work.}
    \label{fig:SEDmodel_illustration}
\end{figure}

Based on these two assumptions, our final model can be mathematically described as follows:

\begin{equation}
    \label{eq:SEDfit}
    f^{\rm total}_{i} = f^{\rm AGN}_{i} + C_{i}\cdot (f^{\rm stars}_{i} + f^{\rm SF}_{i})
\end{equation}

\noindent where $f_i$ are model fluxes in filter $i$, and $C$ is a filter-specific multiplicative constant. The value of $C$ is unity for filters measuring integrated flux (CANDELS and HELP filters), and is determined by the method described below for \textit{JWST} filters that measure the central flux. For clarity, Figure \ref{fig:SEDmodel_illustration} illustrates the contributions of various components in the simultaneous fit.

To determine the relative galaxy emission $C$ in the central SED, we first fit the three component model only to the integrated SED. While most AGN parameters are unconstrained here, this fit still gives an estimate of the relative contribution of the AGN with respect to the host galaxy. The SEDs of galaxy stellar populations peak at approximately $1\;\mu m$ in the rest-frame and the contrast of a galaxy's stellar light to AGN is highest at these wavelengths. Building on this, we estimate the ratio of the AGN model flux to the total model flux at $1\;\mu m$ rest-frame for the best-fit SED of each source. We then classify a source as ``AGN-dominated'' if this ratio is $> 0.1$, and ``galaxy-dominated'' if it is $< 0.1$.

For galaxy-dominated systems, we directly measure the host galaxy contribution in the ``central SED'' by taking the ratio of the measured central (\textit{JWST}) flux to the measured integrated flux at rest-wavelength $1\;\mu m$\footnote{This is done by interpolating between the two photometric points on either side of $1\;\mu m$.}.
For AGN-dominated sources, the host galaxy contribution in the central SED is minimal, but cannot be accurately ascertained from the \textit{JWST} imaging. Therefore, we set the scaling constant $C=0.1$, corresponding to an upper limit to the galaxy emission of 10\% in the central aperture, at the wavelength where the galaxy is brightest. At other wavelengths, the AGN dominance will be even higher, so our choice of imposing an upper value to the galaxy light does not affect the AGN model fit significantly.

\subsection{SED fitting}
\label{subsec:sed_fitting_method}
The composite model described in the previous section has a large number of free parameters, with complex degeneracies between parameters. Therefore we make use of Bayesian inference methods that utilise reasonable prior knowledge to alleviate degeneracies when reasonably motivated from independent population studies. This method not only finds the best-fit parameter values but also provides a realistic estimate of uncertainties along with a complete posterior distribution for each parameter. We make use of these posterior distributions later to assess the constraints on each parameter (Section \ref{subsec:trendwithxray}). 

We have used the Bayesian SED fitting package FortesFit\footnote{Version 2.0.0: https://pypi.org/project/fortesfit/2.0.0} \citep{fortesfit} to perform all the fits in this work. This code provides a framework to incorporate the various SED models described before, generate model photometry for the suite of filter bands we have used, place priors on all parameters, and run the fits using the nested sampling algorithm \citep{skilling04} implemented through Multinest \citep{multinest_feroz08}.

Our general approach for choosing the priors was to use well-established independent relationships between various physical parameters \citep[e.g., the X-ray luminosity and the UV luminosity of AGN; ][]{lusso16_XrayUV}), while keeping uninformative priors for the parameters (e.g., opening angle of the torus) that could not be inferred reliably from literature studies or independent physical constraints. Using these general rules, the specific priors on each of the model parameters are listed below:

\begin{itemize}
    \item Stellar mass (\logMstar): A uniform prior between 8 to 13, in units of $\log$ solar masses. This captures the range between the faintest galaxies detectable in the EGS and the most massive galaxies known in the Universe.
    
    \item Reddening for attenuation of starlight (\EBVstar): A half Gaussian prior with a standard deviation of $\sigma = 0.2$ mag and maximum allowed value of 2.0 mag. This is informed from extinction studies of a sample of $\sim$ 90,000 local galaxies from SDSS \citep{attenuationlocal}.
    
    \item Logarithm of accretion disc luminosity (log(\Laccdisc)): A Gaussian distribution with a mean calculated from \Lxray\;using the relationship between 2 keV luminosity and 2500 \angstrom\ luminosity from \cite{lusso16_XrayUV}. The standard deviation of this distribution was 0.3 dex, which accounts for intrinsic scatter in the linear correlation between luminosities in logarithmic space.
    
    \item Additional extinction along the line-of-sight to the AGN (\EBVagn): A half Gaussian prior centred at 0 and $\sigma = 1.0$. This accounts for the mild extinction that the UV-optical emission from the AGN may experience as it passes through dust in the host galaxy. Red quasars from the Sloan Digital Sky Survey (SDSS) typically experience \EBVagn$<0.1$ \citep{Richards2003}, but we choose a wider distribution to account for the fact that the SDSS is incomplete to the more heavily reddened quasars \citep[e.g.,][]{Fawcett2024}. 
    
    \item $t$, $p$, $q$, $oa$, $R$: A uniform prior over the full range allowed by SKIRTOR.
    
    \item $i$: A uniform prior between 0$^\circ$ and 80$^\circ$ degree. We have excluded 90$^\circ$ as the accretion disc luminosity goes to zero at an inclination of 90$^\circ$ as a result of the illumination law used in SKIRTOR. This prevents us from determining an accurate scaling for \Laccdisc\ from the intrinsic UV luminosity that remains self-consistent with the torus model.
    
    \item 8-1000 \mcron\; IR luminosity of star-formation heated dust (\Lsf): A uniform prior between $42$--$47$ in $\log$ \ergs. This corresponds to dust emission from the full range of star formation rates seen in galaxies that may be detectable at intermediate and high redshifts \citep[e.g.,][]{Magnelli2009}.
    
    \item $\alpha_{SF}$: A Gaussian distribution with a mean of 2.0 and $\sigma = 1.0$. This is informed from local and high redshift AGN studies where $\alpha$ can be constrained \citep{daleHelou02, Dale14, rosario16}.
\end{itemize}

Fits to broadband SEDs can be heavily driven by a small number of points if there are erroneous sharp features or unreliable error estimates on the measured photometry. We employ a few different measures to minimise such issues. First, we impose a minimum error of 10\% on all photometric measurements to avoid a few points with underestimated errors from heavily affecting the fit. Second, we note that HELP photometry is reliant on the deblending of low-resolution \textit{Herschel} images, and some of the resultant deblends have been assessed as being unreliable by the HELP photometric algorithms. We de-weight these measurements by an arbitrary factor of 3 compared to other photometric measurements in which deblending errors are not as pervasive an issue, ensuring that they contribute to the fit but only if reliably deblended \textit{Herschel} measurements do not exist. Third, if the error on the flux is greater than the flux itself then we treat those measurements as upper limits with a value that is the sum of flux and error. Finally, we disregard any photometric measurements that were both identified as limits and assessed as unreliable by the HELP algorithm.

\subsection{Example fits}
\label{subsec:example_fits}
\begin{figure*}
    \centering
	\includegraphics[width=\textwidth]{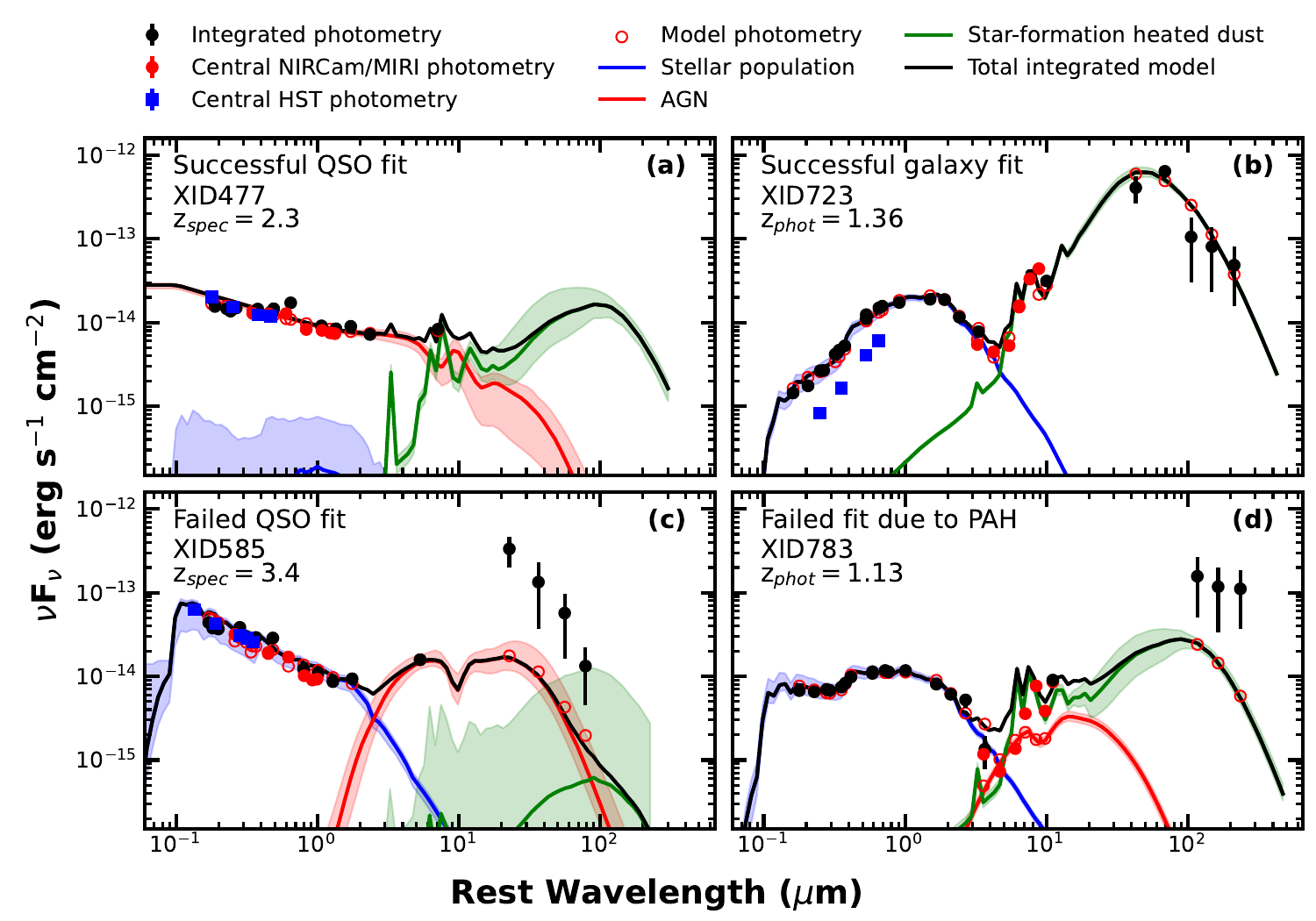}
    \caption{Illustrative examples of some of the SED fits in this study. The source XID, redshift, and the reason for inclusion is given in the top left corner of each subplot. The filled black circles are the ``integrated SED'', filled red circles are central \textit{JWST} SED, and the filled blue squares are central \textit{HST} SED. The open red circles are best-fit model SEDs which the observations should be compared to. The blue, red, and green solid lines show the template SEDs from the original model grid, and the shaded region shows the scatter around best-fit model derived using 100 independent samples from the joint posterior distribution.}
    \label{fig:example_fits}
\end{figure*}

We visually examined the all the fits to understand the capabilities and limitations of our fitting approach. In this process, we identified two galaxies (XID391 and XID506) which showed strong, unphysical discontinuities in their integrated SEDs, likely driven by source mismatches between images with different spatial resolutions. As the rectification of the archival photometry is beyond the scope of our work, we chose to exclude these sources. Their exclusion yields the final sample of 88 AGN, as noted in Section \ref{man_filt}. 

In this section, we visualise the SED fits for some of the sources in the final sample, and discuss the successes and limitations of our SED fitting approach. Figure \ref{fig:example_fits} shows the fits for four different sources selected for specific reasons mentioned in the top left corner. This figure also serves as a guide for interpreting similar figures visualising our SED fits elsewhere in the paper. 

The redshifts for each object are given in the top left corner of each subplot along with its AEGIS-X ID (i.e., XID). The filled circles in the figure are the observed fluxes. The black circles show the integrated SED (obtained from various multi-wavelength catalogues as described in section \ref{subsub:integratedSED}); the red circles show the central SED from \textit{JWST} (measured using the method described in section \ref{subsub:centralSED}), and the blue squares show the central SED from \textit{HST} (also described in section \ref{subsub:centralSED}). The open red circles show the best-fit model photometry; some of these open points are completely hidden behind the solid points, but each observed flux has an associated model flux, except \textit{HST} fluxes that were not used in the fitting. 

The solid lines represent the model SED on the original parameter grid closest to the best-fit model: black, blue, red, and green lines represent total, stellar population, AGN, and galaxy dust models. Because the full model SEDs are only defined by the various model libraries at specific values of the parameters, these visualisations of the SED models are necessarily snapped to the nearest grid points. However, the likelihood evaluation while fitting relies on the interpolation of the model photometry grids, which is more exact. While the lines in these figures serve as a guide to the approximate best-fit model SEDs, we caution the reader that a true assessment of the quality of fit should be done by comparing filled points and open points. Finally, the shaded region shows the scatter around the approximate best-fit model SED derived from the ensemble of models obtained at 100 independent samples from the joint posterior distribution. 

Each of the panels in Figure \ref{fig:example_fits} shows the fit for an object that represents a particular behaviour of our fitting approach. Figure \ref{fig:example_fits}(a) shows a successful fit for a quasi-stellar object (QSO) with spectroscopic $z=2.317$. This is an example of an AGN-dominated system, evident from the much weaker stellar component (blue shaded region) compared to the AGN component (red line) in the optical and NIR region. \citetalias{Nandra15_AEGISX} identified this source independently as a QSO from the characterisation of its SED. Its appearance as a point source in the \textit{JWST} images also supports its QSO nature. Careful examination reveals a excess over the best-fit model near $0.6-0.7 \mu m$, which can be reasonably attribute to an unmodelled H$\alpha$ line. This source is an example of a subset of about 12 such sources likely to be QSOs. They are all fitted with a similar SED model and have low values of the covering fraction.

Figure \ref{fig:example_fits}(b) shows a successful fit for a galaxy-dominated system with no identifiable AGN component (no red line). \textit{JWST} photometry (solid red circles) gives us the power to observe the transition between dust emission (green line) and stellar emission (blue line) by finely sampling the rest-frame NIR. This could not be done with archival data, as this spectral region only had a single observation close to rest-frame $10 \mu m$ from \textit{Spitzer}/MIPS. An excellent fit to the data also shows that our approach is able to robustly characterise systems with little AGN contribution as well.

Figure \ref{fig:example_fits}(c) shows a failed fit where an SED that appears to be a QSO is fit by a galaxy model with a very young stellar population. We suspect this could be because the fixed shape of the unobscured AGN spectrum in SKIRTOR, which is unable to models QSOs with steeper than typical UV-optical spectral slopes. There is only one other such source in our final sample and we keep both of them in our analysis as they are, instead of manually forcing a quasar fit, to capture any systematic biases in our SED modelling.

Figure \ref{fig:example_fits}(d) shows an example of a failed SED fit where the best-fit model SED (open red circles) lie significantly below the observed \textit{JWST} photometry (solid red circles). This is likely due to inadequate modelling of the complex set of PAH features around 7-12 \mcron, which is leading to an inaccurate normalisation of the AGN model. A full treatment of the variation found among the PAH features in galaxies is beyond the scope of this work, partially because the MIRI coverage is sparse for most of our sample. We have verified that this is the only source suffering from this issue. Therefore we consider it to be a limitation of our fitting approach and keep the source in our sample, fitted as is.

%%%%%%%%%%%%%%%%%%%%%%%%%%%%%%%%%%%%%%%%%%%%%%%%%%%%%%%%%%%%%%%%%% 
%                           Results
%%%%%%%%%%%%%%%%%%%%%%%%%%%%%%%%%%%%%%%%%%%%%%%%%%%%%%%%%%%%%%%%%% 
\section{Results and Discussion}
\label{sec:results}
The main advantage of the SED fitting framework described in this work is the ability to separate the IR emission of the AGN torus from the large amount of host-galaxy contamination using high-resolution \textit{JWST} imaging. This purer estimate of the AGN emission, when fit simultaneously with the integrated emission, gives more robust constraints on both AGN and galaxy parameters. We examine the constraining power of this approach for various parameters in Section \ref{subsec:trendwithxray}. We define a subsample of 66 sources where we can reliably constrain an important AGN parameter -- the covering fraction ($CF$). We compare the $CF$ distribution of our sample with that of a local AGN sample in Section \ref{subsec:cf_compare} to search for signs of evolution. We also discuss the $CF$ distribution in the context of X-ray obscuration and the AGN type (Type-1 vs. Type-2). In Section \ref{sec:moreparams}, we discuss the distributions of other AGN and galaxy parameters which further validate our SED fitting approach. In Section \ref{sub:resolvedJWST_result}, we show the improvements in constraining power brought about by the use of high-resolution \textit{JWST} imaging through both specific examples and population level statistics. Finally, in Section \ref{sub:improvements}, we discuss some caveats along with step taken to address them. 

\subsection{SED fits: validation, trends, and population-level statistics}
\label{subsec:trendwithxray}

The X-ray luminosity (\Lxray) is an independent measure of the AGN power in our sources and is therefore expected to correlate with the contribution of the AGN emission relative to the host galaxy in the integrated SEDs. This relative AGN contribution in turn relates to the ability to constrain the geometrical parameters of the torus using IR observations, which is the primary focus of this study. In this section, we examine the trends between various model parameters and \Lxray. 

\begin{figure*}
    \centering
	\includegraphics[width=\textwidth]{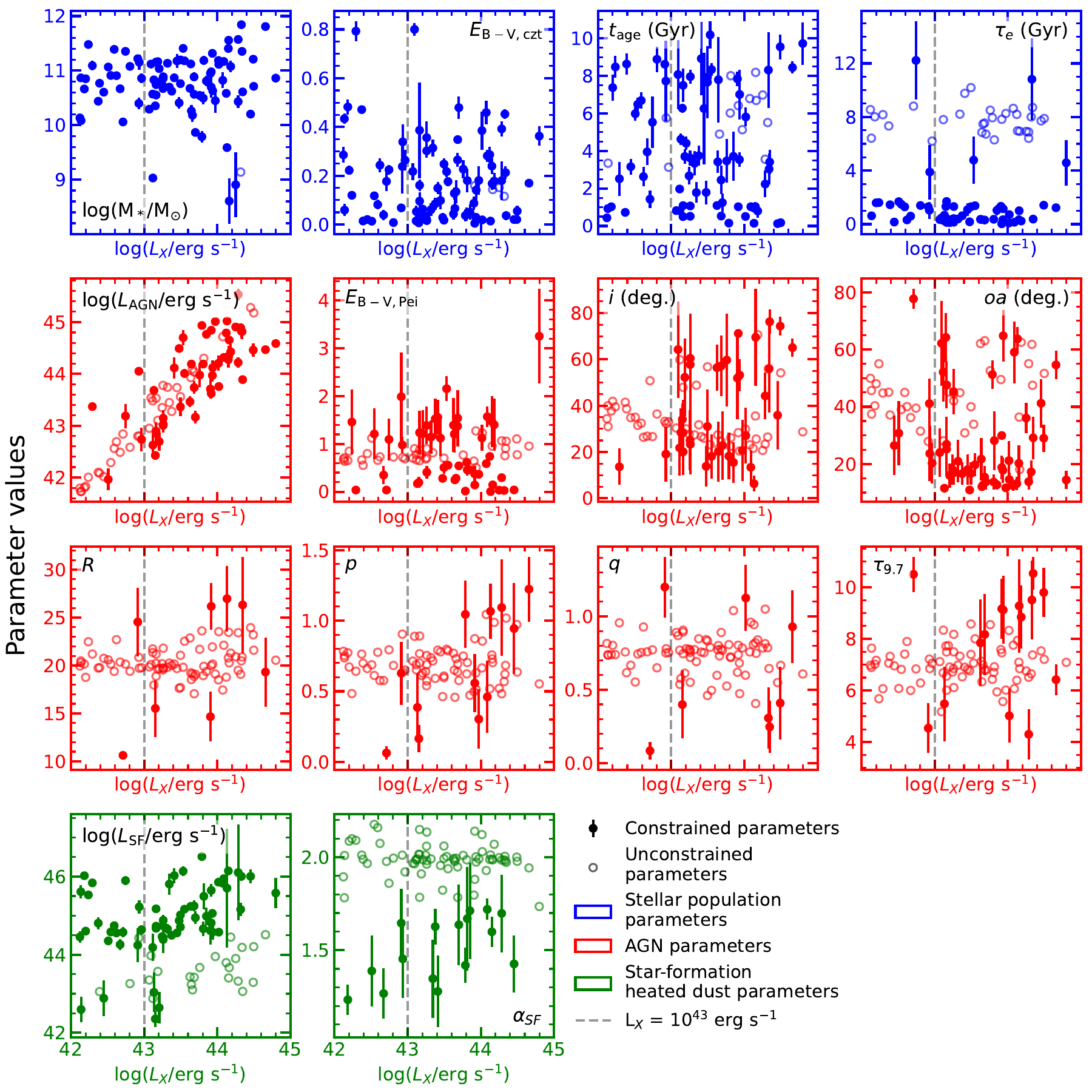}
    \caption{The distribution of the best-fit values (defined as the median of posterior distribution) for all free model parameters for 88 sources plotted against the $2-10$ keV X-ray luminosity (\Lxray) on the horizontal axis. The filled circles correspond to the sources where the given parameter was constrained by SED fitting, while the open circles correspond to the sources where that parameter was unconstrained (see Section \ref{subsec:trendwithxray} for more details). The vertical dashed line shows the \Lxray$\geq 10^{43}$\;\ergs\; cut used to define a subsample where AGN parameters can be reliably constrained. Finally, the colours of the frame show the SED model component to which the given parameter belongs.}
    \label{fig:population_stats}
\end{figure*}

For the subsequent analysis, we first outline a method to determine whether the SED fitting process has effectively constrained a specific model parameter for a given source. We define a parameter as ``constrained'' if its posterior distribution differs significantly from its prior. This is to say that the SED fitting has led to updating of the prior knowledge. We measure the dissimilarity of the two distributions by calculating the Kullback-Leibler Divergence (KLD); low KLD values indicate little change (i.e. similar distributions), while high KLD values suggest that the posterior has diverged from the prior \citep{KLD}. We determine a threshold KLD for individual parameters by inspecting plots of KLD versus the $1\sigma$ uncertainty across all sources (see Appendix \ref{app:klderror} for details). We will comment on the impact of this constrained/unconstrained distinction on our results as we proceed.

Using aforementioned definition of constrained and unconstrained parameters, Figure \ref{fig:population_stats} presents the best-fit values\footnote{Defined as the median of the posterior distribution.} of the 14 free model parameters for the 88 sources in our sample plotted against \Lxray. The filled circles correspond to the sources where the given parameter is constrained, and the open circles indicate sources where the parameter remains unconstrained. We find no obvious correlations between model parameters and \Lxray\;except for the correlation with \Laccdisc, which is partially set by the prior used on the accretion disc luminosity during the SED fits. Although we will use the full posterior distributions for population-level analysis, Figure \ref{fig:population_stats} provides a visual summary of the overall fit quality and the effectiveness of our SED fitting method in constraining model parameters. It also serves as a reference for identifying subsets of our sample where AGN parameters are reliably constrained. We also investigated similar trends with redshift, another key variable, but observed no significant trends across the redshift range covered by this study.
 
The red panels in Figure \ref{fig:population_stats} show the trends of the best-fit AGN parameters against \Lxray. We see that the fraction of sources with constrained AGN parameters falls drastically below \Lxray\;$ = 10^{43}$ \ergs, evident by many more open circles compared to filled circles left of the black dashed line (see specifically log(\Laccdisc), $i$, and $oa$ panels). This is likely because these low-luminosity AGN produce weaker IR emission that makes it difficult to distinguish from the brighter host-galaxy emission at these wavelengths, which severely limits our ability to constrain the shape of the AGN SED. A similar correlation between constraining power and the AGN contribution to the overall SED is reported in other studies \citep[e.g.,][]{Ichikawa19}. Due to this, we restrict our statistical analysis to the subset of sources with \Lxray\;$ \geq 10^{43}$ \ergs, the region to the right of the black-dashed line in Figure \ref{fig:population_stats}. This cut reduces the sample size from 88 to 66.

Figure \ref{fig:population_stats} also shows that certain AGN parameters, such as those related to dust distribution and density in the torus ($p$, $q$, \Rratio, and \opticaldepth), are almost never constrained by SED fitting. It is difficult to constrain these parameters because they produce minimal variation in the broadband SED shapes despite having significant influence on narrower spectral features like silicates and PAH lines. Indeed, these parameters are rarely constrained even in local studies with much higher quality broadband SEDs \citep[see ][]{yang2020, SKIRTOR_local}. Nonetheless, we opted to keep these parameters free in our SED fitting to allow for the full range of SED shapes predicted by the models. This approach provides more realistic posterior distributions on the parameters that can be constrained (such as the opening angle), and therefore allow a more accurate treatment of the uncertainties on these parameters.

While the detailed modelling of galaxy light is not the primary focus of this work, we utilize these parameters for both qualitative and quantitative validation of our fitting method by comparing them to well-established galaxy properties in the EGS, as discussed in greater detail in Section \ref{sec:moreparams}. However, just from Figure \ref{fig:population_stats} we see that the relative number of sources with constrained and unconstrained galaxy parameters (shown in the blue and green panels) is uniform across the X-axis, and, consequently, not dependent on \Lxray. The quality of fit to the galaxy emission is largely independent of the luminosity of the AGN.

The last panel of Figure \ref{fig:population_stats} shows the best-fit values of $\alpha_{SF}$, a parameter that is sensitive to the relative amounts of star-forming dust at different temperatures. Interestingly, all the sources with constrained $\alpha_{SF}$ (filled circles) have low values ($\alpha_{SF} < 2$), corresponding to a large fraction of hotter dust. An examination of the fits of these objects reveals that all of them have high FIR luminosity from star-formation (\Lsf\;$ > 10^{44}$ \ergs). This is consistent with previous studies that have shown that starburst galaxies, with strong FIR emission and therefore better constraints from \textit{Herschel} photometry, tend to have warmer SEDs with a larger proportion of high temperature dust \citep[e.g.,][]{daleHelou02}.

Our final goal in this work is to study the ensemble or population-level properties of our sample, particularly those of the torus. However, the population statistics inferred only from best-fit models will be misleading because they do not incorporate the uncertainties in parameter estimates of individual sources. We tackle this issue by using a bootstrap approach that utilises the full posterior distributions of the parameters for each source. 

The output of the fits are a set of samples from the full 14-dimensional joint posterior distribution, supplied by MultiNest, with around 2,000--6,000 samples per object. To construct a single instance of a population-level distribution, we draw one sample from the posterior distribution of each of the 66 AGN, and combine these together to extract a distribution for each parameter. We repeat this process 5,000 times with replacement, yielding 5,000 instances of such population-wide distributions. Each of these draws is a different, independent realisation of the parameter distributions of the entire sample. 

The parameter distributions from each draw are binned in the same fashion to generate histograms. From the full set of draws, the median value of the resultant histograms, as well as the 16-th and 84-th percentile ranges, give us the final population-level distributions along with statistical uncertainties. This approach provides a robust representation of these distributions, accounting for both uncertainties in individual parameter estimates and variations across the sample. The resulting distributions are discussed below, starting, in the next section, with the key parameter, the covering fraction ($CF$) of the torus, followed by selected galaxy parameters in Section \ref{sec:moreparams}.

\subsection{The torus Covering Fraction ($CF$)}
\label{subsec:cf_compare}
$CF$ is defined as the proportion of the sky's solid angle subtended by the obscuring torus as seen from the central SMBH. It is given by $CF = \sin(oa)$, where $oa$ is the half opening angle of the torus measured from the equatorial plane to the upper edge of the torus, as defined in SKIRTOR. This parameter is a measure of the ability of the torus to obscure the central engine, and plays an important descriptive role in various theories of torus formation and evolution. Given its importance, we examine $CF$ in detail in this section and compare it with the similar AGN populations in the local universe.

\begin{figure*}
    \centering
	\includegraphics[width=0.8\textwidth]{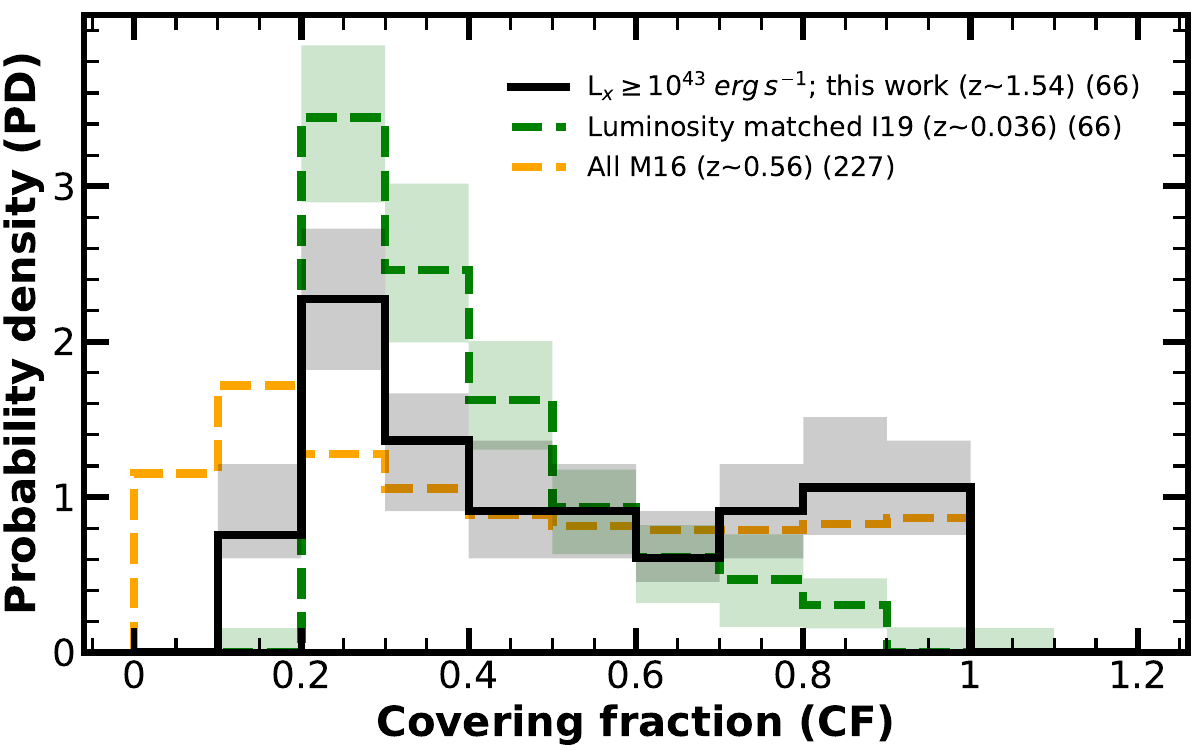}
    \caption{The covering fraction ($CF$) of various samples of AGN. Solid black line shows the median $CF$ distribution of all 66 sources in our subsample with \Lxray\;$\geq 10^{43}$ \ergs. The shaded region shows the uncertainty on the distribution calculated using the procedure described in Section \ref{subsec:trendwithxray}. The green dashed line shows the $CF$ distribution for the local \citep[median $z \sim 0.036$;][]{Ichikawa19}, luminosity matched AGN population with the green shaded region showing the $1\sigma$ spread due to multiple iterations of luminosity matching. The orange dashed line show the intermediate-redshift \citep[$z \sim 0.56$;][]{MateoslocalCF} AGN population.}
    \label{fig:CFcompare_others}
\end{figure*}

The distribution of median $CF$ of the population of 66 sources in our sample, those with \Lxray\;$\geq 10^{43}$ \ergs, is plotted by the black line in Figure \ref{fig:CFcompare_others}. The statistical uncertainty on the distribution is shown by the grey shaded regions, determined by the bootstrap approach laid out in Section \ref{subsec:trendwithxray}. The distribution is broad with a peak at $CF \approx 0.25$. A similar peak is seen for other X-ray selected AGN (see below), but not as clearly for AGN selected by other means, such as local Seyferts which were identified based on their emission line properties \citep{RA11, GB19}. The $CF$ distribution we find is likely to be shaped by selection effects: our sample is incomplete towards the most strongly X-ray obscured AGN, those with intervening high column densities towards their nuclei, which are also likely to have tori with large covering factors. 

To contextualise our findings, we compare our $CF$ distribution to other X-ray selected AGN populations at low and intermediate redshifts.

\subsubsection{Comparison with local X-ray selected AGN} 
\label{subsubsec:localcompare}
We start with the local AGN sample from \cite{Ichikawa19} (I19 hereafter) containing 587 sources detected in the 70-month Swift/BAT survey based on their emission in the $14-150$ keV energy band\footnote{Note that the Chandra energy bands used to select our AGN sample corresponds to $\sim 10-30$ keV at $z \sim 1$--$2$, capturing the ``Compton hump'' feature from the torus which is also the dominant emission in $14-150$ keV energy band.}. The median redshift for this sample is $\approx 0.036$. We construct luminosity matched sub-samples from this parent sample by pairing each of our sources with exactly one source from \citetalias{Ichikawa19}, without repetition, within 0.2 dex in bolometric luminosity (\Lbol)\footnote{The bolometric luminosity in both cases was calculated using luminosity-dependent bolometric correction given in Equation 21 of \cite{marconi04}.}. This careful matching ensures that any observed difference in the $CF$ distribution is not biased by the luminosity-based variation, as required by the receding torus model and reported in numerous studies \citep[e.g.][]{Lawrence91, ueda03, toba2014, ichikawa17}. The distribution thus obtained is shown with green line in Figure \ref{fig:CFcompare_others}, and the green shaded region represents the scatter due to multiple iterations of luminosity matching process.

\citetalias{Ichikawa19} calculated the $CF$ of each source using its relationship with $L_{\rm torus}/L_{\rm AGN}$ based on SKIRTOR (Table 1 of \cite{skirtor16}). They determined the ratio $L_{\rm torus}/L_{\rm AGN}$ by fitting the MIR-FIR ($\sim 10-500$ \mcron) SED using empirical galaxy and AGN templates that are different to the ones used in our work. However, since the $CF$ reported in \citetalias{Ichikawa19} is ultimately calibrated to SKIRTOR, we can directly and self-consistently compare our $CF$ distribution to theirs. Doing this, we find that both these distributions show a consistent peak at low $CF$. However, the \citetalias{Ichikawa19} $CF$ distribution drops rapidly at high $CF$ ($> 0.6$) while that from our work is largely flat beyond 0.4. Visually, the differences are greater than the statistical spread in our distribution, as the green line differs from the black line beyond the shaded regions in Figure \ref{fig:CFcompare_others}, suggesting a possible evolution in $CF$.

We perform a two-sample KS test to statistically verify this result. To do this, we select a random realisation of each of the two populations (i.e. a distribution from the shown spread) and calculate the KS statistic for 1000 such pairs of realisations. The median p-value of the KS test between \citetalias{Ichikawa19} and our sample (dashed green line and solid back line in Figure \ref{fig:CFcompare_others} respectively) is 0.066 with median KS statistic value of 0.2273. This means that we can not reject the null hypothesis that these two samples are derived from the same population with generally accepted $> 95\%$ certainty even though they look qualitatively different.

Our analysis also suggests that the decisions made during the SED fitting process (specifically, choice of priors and fitting technique) can sometimes also appear as signs of evolution. We discuss this cautionary analysis in Appendix \ref{app:pointest}.

\subsubsection{Comparison with an intermediate redshift sample}
The dashed orange line in Figure \ref{fig:CFcompare_others} shows the $CF$ distribution of an intermediate redshift AGN sample from \cite{MateoslocalCF} (M16 hereafter). These AGN were drawn from the Bright Ultra-hard XMM-Newton Survey (BUXS) based on their $4.5-10$ keV detections. They range up to $z = 1.7$, with a large fraction below $z = 1$, and have $10^{42} <$ \Lxray $< 10^{46}$ \ergs. \citetalias{MateoslocalCF} decompose the optical-MIR SED using SEABAS \citep{seabas}, and fit the torus emission with the 6-parameter CLUMPY torus model \citep{Nenkova08} using the BayesCLUMPY SED fitting code \citep{bayesclumpy}. Unlike the geometrical $CF$ that we use, the $CF$ in \citetalias{MateoslocalCF} was calculated using the radiation escape probabilities for the torus geometries used in their SED fits, taking into account the full posterior distributions. The sample also includes sources with unconstrained $CF$ even though the authors do not make an explicit distinction.

We see that $CF$ distribution from \citetalias{MateoslocalCF} (orange dashed line) is consistent with the $CF$ distribution of our population (given by solid black line) within uncertainties except at low $CF$. This apparent deviation is due to the difference in $CF$ definitions between the two torus model families. For example, unlike CLUMPY, SKIRTOR's $CF$ has a lower bound of $\approx0.17$ set by a minimum allowed angular thickness of the torus of $oa < 10^{\circ}$. This is also why the KS test for these two distributions returns an extremely low p-value ($\approx 0.003$), but is an inappropriate measure to compare these two populations with different $CF$ definitions. Nevertheless, the broad trends in these two distributions display an enhancement in sources with low $CF$ as well as flat tail towards higher $CF$. We conclude that they are qualitatively consistent with each other barring the model differences.

Given our comparisons with AGN samples in local (\citetalias{Ichikawa19}) as well as intermediate (\citetalias{MateoslocalCF}) redshifts we conclude that there is minimal evidence for $CF$ evolution out to $z\sim2$.

\subsubsection{Comparison with redshift evolution from other studies}
\begin{figure}
    \centering
	\includegraphics[width=\columnwidth]{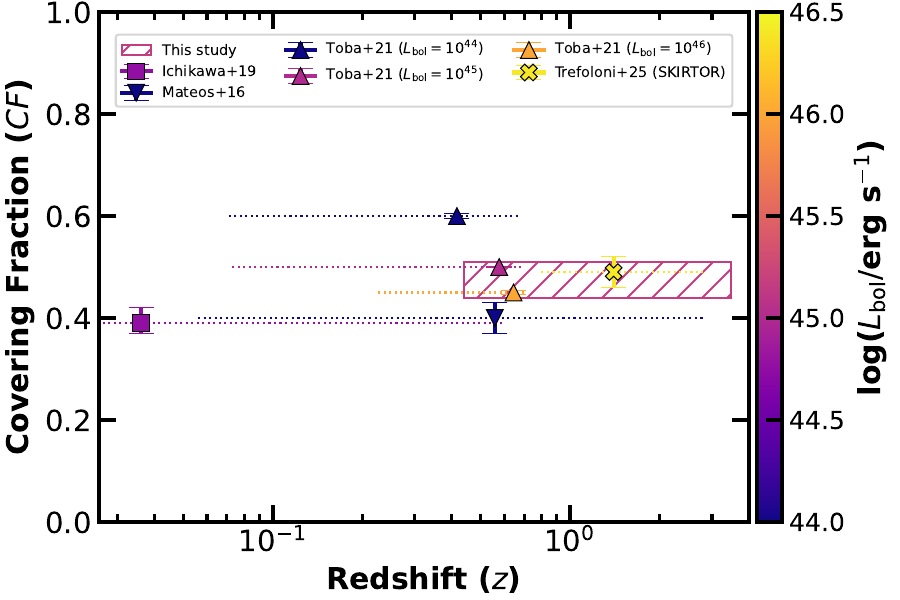}
    \caption{The median value and standard error (note this is not same as standard deviation) on the $CF$ in different populations in literature. All $CF$ are obtained using different methods. The colour of the point shows the median (or mean in case of \protect\cite{MateoslocalCF} and \protect\cite{trefoloni25} as median is unavailable) bolometric luminosity, and the horizontal dotted line shows the full redshift range of that sample. The hatched rectangle gives the redshift range and standard error covered by our sample, with colour of the rectangle corresponding to the median bolometric luminosity.}
    \label{fig:CFcompare_other}
\end{figure}

Figure \ref{fig:CFcompare_other} compares the median $CF$ obtained in this work with those from other SED-focused studies spanning a wide range of redshifts. Each of these studies differ in their methodology and selection, but the figure still provides a comparative view of $CF$ over redshift. The figure shows that our study probes significantly lower luminosity AGN compared to previous studies around cosmic noon \citep[e.g.][]{trefoloni25}. We do not see any significant trend in the $CF$ with redshift between these samples. However, we do see a trend between $CF$ and bolometric luminosity in the \cite{toba21} sample where the $CF$ is measured self-consistently between different luminosity bins. This is expected from the receding torus model as discussed in \cite{toba21}.

This result would suggest a lack of evolution in $CF$, but we caution against this simplistic view. This representation fails to capture the full spread of $CF$ values as well as the shape of underlying $CF$ distribution of the populations. As show in Figure \ref{fig:CFcompare_others}, both these could have non-trivial effects on the comparison. Moreover, the difference in selection, $CF$ inference, and large redshift ranges could also dilute some signs of evolution. We address these points in the upcoming controlled study with a larger AGN sample.

\subsubsection{Covering fraction distributions split by AGN obscuration}
\label{subsub:CF_obscuration}
\begin{figure}
    \centering
	\includegraphics[width=\columnwidth]{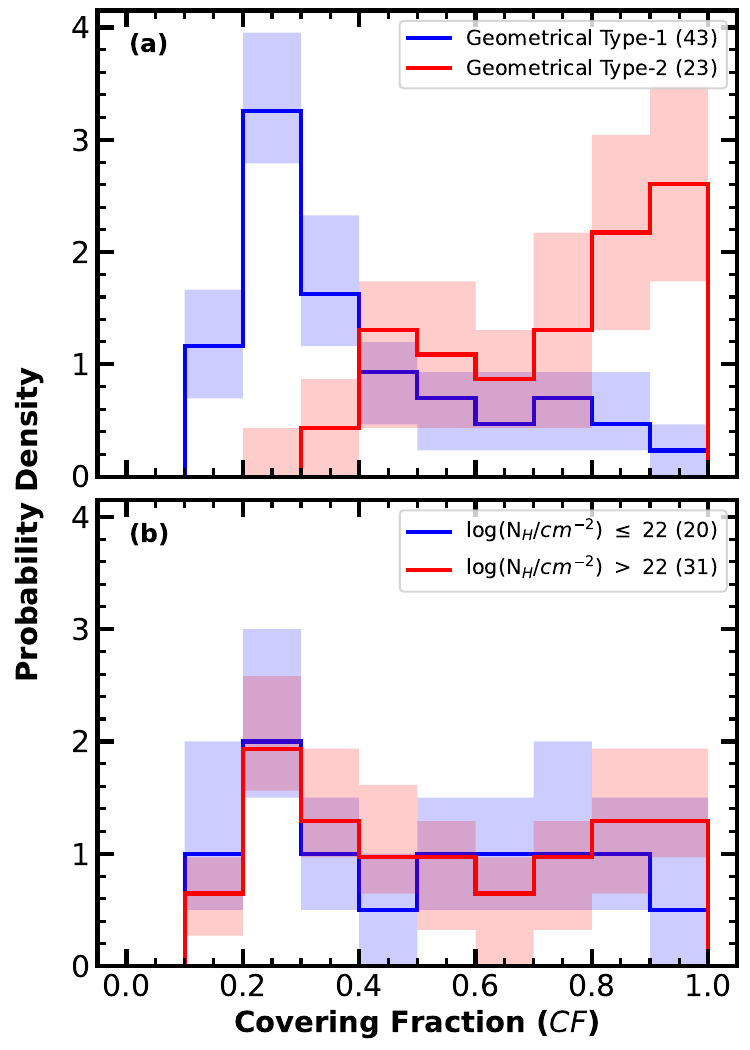}
    \caption{Covering fraction ($CF$) distributions for different subsets of the 66 sources with \Lxray\;$\geq 10^{43}$ \ergs. Panel (a) shows the $CF$ distribution for Type-1 (blue) and Type-2 (red) AGN as defined by the geometry of the torus (refer to Section \ref{subsub:CF_obscuration}), and panel (b) shows the $CF$ distribution for two subsets based on \NH. The numbers in the brackets in legend indicate the number of sources in that subset.}
    \label{fig:CFcompare_type}
\end{figure}

A number of previous studies have shown intrinsic differences in the $CF$ distributions of Type-1 and Type-2 AGN, the spectral classes that correspond to unobscured and obscured AGN, respectively. Type-1 are biased towards low $CF$, while Type-2 typically have high $CF$; a simple consequence of more sight-lines being obscured by the high $CF$ torus for randomly distributed inclinations as noted in \cite{type_CF}. These studies \citep[e.g.,][]{RA11} have asserted that the differences in intrinsic torus geometry are fundamental elements of the Type 1/2 dichotomy among AGN.

In absence of detailed UV/optical spectroscopy for our sources, their Type-1/Type-2 nature is a-priori unknown. We therefore define geometrical Type-1 or Type-2 AGN based on whether our fits predict a clear view of the accretion disc along the line-of-sight. We determine this by calculating the median value of the sum of the torus half opening angle ($oa$) and torus inclination angle ($i$) from the joint posterior distributions of these parameters. Recall that, as per the definition of $oa$ and $i$, if $i > 90^{\circ} - oa$, then the line-of-sight intersects the torus and we see an obscured AGN. In this fashion, we can classify objects with median  $oa + i < 90^\circ$ as Type-1 and those with median sum $oa + i > 90^\circ$ as Type-2. We caution again that such a geometrical Type-1 and Type-2 split is not a perfect model-independent classification, but it is still useful for studying differences in the absence of spectroscopic hallmarks of the AGN Type. Figure \ref{fig:CFcompare_type}(a) shows the $CF$ distributions of our Type-1 and Type-2 AGN, classified as above. We clearly see that Type-1 AGN (blue histogram) have a lower average $CF$ than Type-2 AGN (red histogram). While the dichotomy in the $CF$ distribution exists for our Type-1 and Type-2 sources, we still find a substantial fraction of Type-1 sources with high $CF$, as well as a range of $CF$ for Type-2 sources, visible in the long tails in both the distributions in Figure \ref{fig:CFcompare_type}(a). 

There are nearly twice as many Type-1 AGN (43) in our sample than Type-2 (23). This is the main reason for the observed peak of low covering fraction in the overall population (solid black line in Figure \ref{fig:CFcompare_others}). X-ray selection is known to be biased towards Type-1 AGN since they are more likely to be unobscured along the line-of-sight, not only providing an unobstructed view of the BLR but also making it easier for the X-rays to escape without absorption by the dense gas in the torus. 

\citetalias{MateoslocalCF} report a similar ratio of Type-1 and Type-2 AGN in their sample, which was also selected using X-rays. We note that the median values of $CF$ in our work are slightly different for those reported in \citetalias{MateoslocalCF}, likely because of the differences in the model ranges for the $CF$, as previously discussed.

Another common method of distinguishing between unobscured and obscured AGN is using the line-of-sight hydrogen column density with a typical cut being $\log(N_H) < 22$ for unobscured and $\log(N_H) \geq 22$ for obscured AGN. We obtained the log(\NH) estimates for 51 out of 66 sources from X-ray spectral fits given in \citetalias{Buchner_AEGISX} and divide them according to this cut. The $CF$ distributions for the two subsets are compared in Figure \ref{fig:CFcompare_type}(b). The population-level $CF$ distributions of the both subsets are remarkably consistent with each other, even though they both constitute different fractions of the overall X-ray selected AGN sample. This is in marked contrast to the dichotomy $CF$ distributions based on optical-IR classifications, indicating that X-ray obscuration and optical obscuration are not strongly correlated. Previous studies in the local AGN population have also found evidence of this \citep[e.g.,][]{Burtscher2016}.

\subsection{Distributions of other physical quantities}
\label{sec:moreparams}
\begin{figure*}
    \centering
	\includegraphics[width=\textwidth]{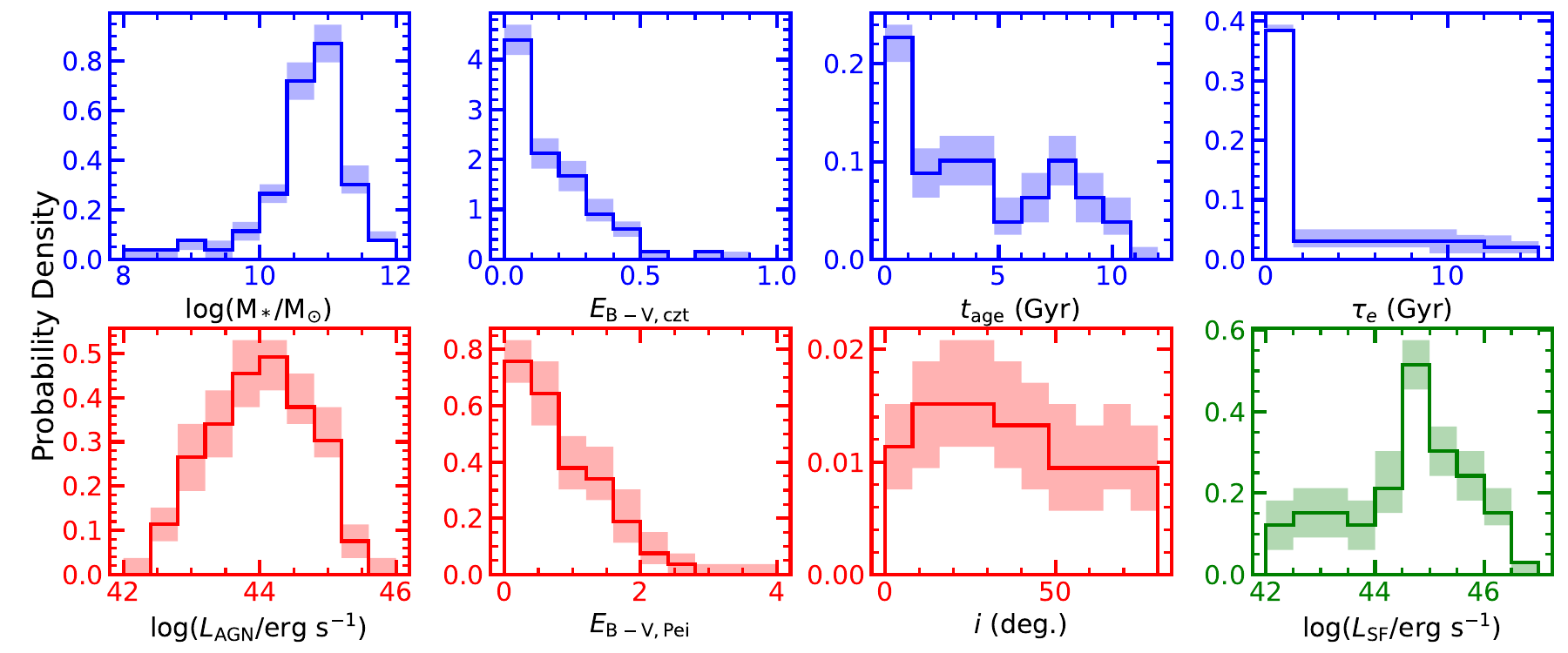}
    \caption{The population-level distributions of a few key parameters, well-discussed in the literature and well-constrained by our fits, of all 66 sources with \Lxray $\geq 10^{43}$\;\ergs. The histograms were constructed using 5,000 realisations of the population by randomly sampling a single chain from the posterior distribution of each source as described in Section \ref{subsec:trendwithxray}. The solid lines and shaded regions show median and $1\sigma$ spread of these realisation respectively. The blue, red, and green colours correspond to free parameters in stellar population, AGN, and dust heated by star formation models respectively.}
    \label{fig:population_histogram}
\end{figure*}

Here we explore the population-level statistics of some of the other galaxy and AGN parameters of the 66 sources with \Lxray\;$\geq 10^{43}$ \ergs, and discuss them in light of the known properties of galaxies and their evolution, as captured by surveys such as CANDELS. Figure \ref{fig:population_histogram} shows the distributions of select parameters that are generally well-known in literature and well-constrained by our fits, generated using the bootstrap method described at the end of Section \ref{subsec:trendwithxray}. 

The host galaxy stellar mass distribution, shown in the first panel of Figure \ref{fig:population_histogram}, peaks around $10^{11} {\rm M}_{\odot}$. This is consistent with other X-ray selected AGN population studies at cosmic noon \citep[e.g.,][]{stellarmass_rosario13, Azadi2017}. This independently validates our approach of fitting the stellar component in our AGN, though, admittedly, stellar mass is not strongly sensitive to the choice of stellar population modelling scheme \citep{Pacifici2023}. 

The distribution of \Lsf\ peaks just below $10^{45}$ \ergs\ with a long tail to lower values. The typical star formation rate of massive galaxies like our AGN hosts at $z \sim 1$--$2$ is $\sim 50$--$100$ M$_{\odot}$/yr \citep{Whitaker2012}, which corresponds to \Lsf\ in the range of $10^{45\textrm{--}45.5}$ \ergs, following standard scaling between IR luminosity and the star formation rate \citep{KennicuttEvans2012}. Therefore, the AGN in our sample are typically slightly weaker star-forming galaxies than those on the star-forming Main Sequence, but also include a number of starbursts (\Lsf$\approx 10^{46}$ \ergs), as well as a number of quenching or quenched galaxies. We caution that the tail towards low values of \Lsf\ is populated by AGN hosts that have only upper limits on their contributions from star formation-heated dust.

Looking at other galaxy properties we find that the attenuation of stars due to the dust in the galaxy is mostly below \EBVstar$< 0.5$ or $A_{\rm V, stars}$ < 2.5 with a small tail. This is consistent with previous studies of a galaxy population at cosmic noon \citep{Av_star_z2}, and validates our choice of prior on \EBVstar, in the sense that the overall population statistics match the prior distribution even though the distributions for individual sources might vary. 

The distributions of the star-formation history parameters demonstrate that most of the AGN hosts have a short star-formation timescale \Tgyr, but display a range of ages. This is expected for massive galaxies, which tend to have formed the bulk of their stars at earlier epochs. Our distributions are completely consistent with earlier estimates of the star-formation histories of X-ray selected AGN hosts at these redshifts \citep[e.g.,][]{stellarmass_rosario13, Azadi2017}.

Turning to AGN properties, the UV accretion disc luminosity ($L_{AGN}$) distribution in Figure \ref{fig:population_histogram} peaks broadly around $10^{44}$ \ergs, equivalent to a 2500 \AA\ absolute AB magnitude of $\approx 21$. These luminosities are fully consistent with those expected from their X-ray luminosities, partly by construction as we apply a prior on $L_{AGN}$ using \Lxray.
The reddening along the line-of-sight (\EBVagn) is a rough power-law with almost no objects showing values $>3$. This distribution is substantially wider than those found among the more luminous QSO population \citep[e.g.,][]{Fawcett2023}, where E(B-V) values $> 1$ are very rare. Because optical extinction through the galaxy is degenerate with the optical extinction through the torus, the population-level distribution of \EBVagn\ that we have obtained probably reflects the conservative prior that we have applied on this parameter. However, our AGN are not pre-selected to be Type-1s and will contain a substantial fraction of systems where dust in the galaxy can influence the observed AGN SED. Therefore, our conservative prior captures the degree of uncertainty we may expect from this poorly-constrained property of our AGN, which can only be properly ascertained by detailed extinction modelling of the rest-frame optical/IR spectra of their hosts.

\subsection{Benefits of resolved \textit{JWST} photometry}
\label{sub:resolvedJWST_result}
\begin{figure*}
    \centering
	\includegraphics[width=0.95\textwidth]{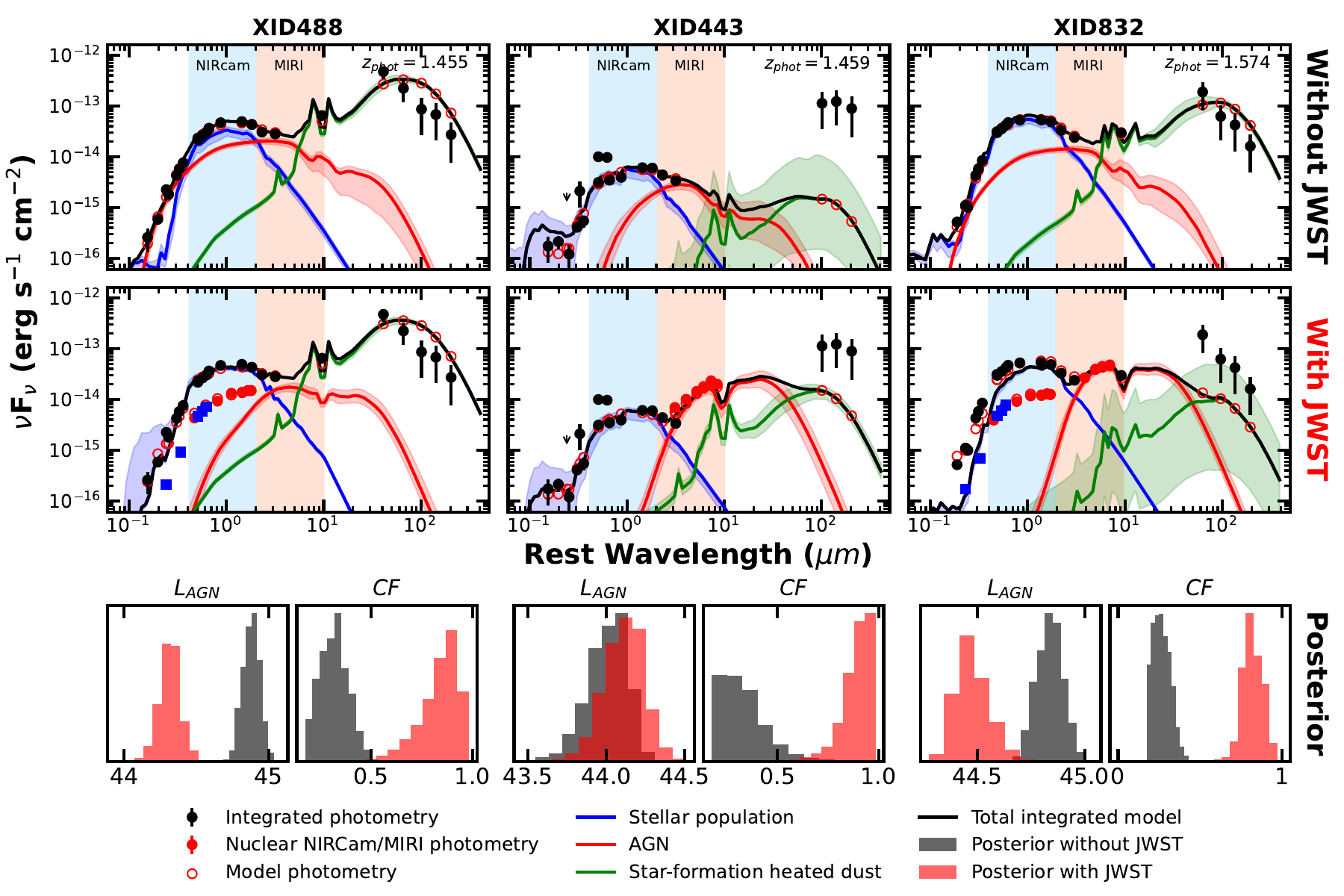}
    \caption{A comparison between the best-fit SEDs with corresponding posterior distributions of \Lxray\; and $CF$; when using central \textit{JWST} SED for SED fitting as compared to when using only the integrated SED. The top row shows the best-fit SED when only the integrated SED (Section \ref{subsub:integratedSED}) is used. The middle row shows the best-fit SED when the nuclear SED from \textit{JWST} (Section \ref{subsub:centralSED}) is also included. The bottom row shows the posterior distribution for these two cases in black and red respectively. The three columns correspond to three different sources selected based on their \textit{JWST} coverage.}
    \label{fig:example_fits2}
\end{figure*}

In Figure \ref{fig:example_fits2}, we demonstrate the impact of high-resolution multi-band \textit{JWST} photometry on the capabilities of SED fitting of AGN hosts. As larger regions of sky garner similar levels of \textit{JWST} coverage, the approach laid out in this work will allow us to explore more diverse AGN populations in similar ways, expanding these valuable analyses to a larger spread in accretion luminosity and in redshift.

The first (left-most) column of Figure \ref{fig:example_fits2} compares two fits for one of our sources, XID488. The top panel shows the fit to only the integrated SED, while the middle panel shows the simultaneous fit to the integrated and central SEDs (the final fits used in the population-level work described above). Both of these are good fits to the data. However, the shape of AGN SED is significantly different between the two. This is reflected in the significantly different posterior distributions of \Laccdisc\; (the accretion disc luminosity) and $CF$ (the covering fraction), both shown in bottom panel. Specifically, \Laccdisc\ decreases and $CF$ increases when the central photometry from \textit{JWST} is included. This is because the central SED (solid red circles) has a shape that is entirely consistent with a galaxy SED, borne out by the sharp 4000 \AA\ break feature seen clearly when the \textit{HST} photometry is included (solid blue squares). For this emission to arise from stars, our fits put an upper bound on the possible reddened UV-optical emission from the AGN, a bound that does not exist when considering the integrated SED alone. Such behaviour is seen in a number of our sources, and, in general, we find that the photometry from NIRCam is particularly useful for constraining the emission from the AGN's accretion disc and its contribution to the SED of the system. 

The second column in Figure \ref{fig:example_fits2} shows another source (XID443), for which only \textit{JWST}/MIRI images exist. From fits to the integrated photometry, this source appears to have a low $CF$ torus (black histogram in the $CF$ posteriors plot shown the bottom panel). However, the unprecedented well-sampled MIR SED from MIRI (solid red circles) is able to isolate the IR emission from the torus which is instead identifiable as one with a high $CF$ (red histogram in the respective posteriors plot). This example highlights the importance of MIRI observations for detecting torus emission, which would have been completely unknown for this source before. Such information is especially important for obscured AGN which have significant torus emission only detectable in the MIR. Even after a fundamental change in our understanding of the torus in this system, \Laccdisc, which is set by the normalisation of the AGN model SED, remains quite similar between the two fits. In general, we find that the MIRI photometry are helpful for constraining the shape parameters of the torus. We note that, in this case, the improved AGN modelling also improves the precision of the galaxy dust modelling in FIR (green line).

The third (right-most) column in Figure \ref{fig:example_fits2} shows the fits for XID832, a source with both NIRCam and MIRI observations. Like before, we see that the NIRCam observations set an upper bound on the amount of UV-optical emission from AGN, and the MIRI observations finely sample the previously unexplored MIR SED dominated by torus emission. These two factors together lead to significantly different \Laccdisc\ and $CF$ posteriors, compared to the fit without the \textit{JWST} SED (bottom right panels). 

Overall, we find that the central photometry from both NIRCam and MIRI is valuable in gaining a better understanding of the AGN as well as its host. The NIRCam observations are mainly important for constraining the unobscured AGN luminosity, which often lead to reduced uncertainty in the host galaxy component as well as the torus parameters. The MIRI observations directly capture the torus emission with minimal host contamination and are more useful for constraining the geometrical parameters of the torus. MIRI observations are also a powerful tool for accurate modelling of the galaxy dust SED alongside the confusion-dominated \textit{Herschel} photometry, as seen in the second and third columns of Figure \ref{fig:example_fits2}. These examples also indicate that the torus parameters (especially $CF$) can be constrained more reliably in the sub-population of AGN where the IR emission from torus is dominant over the galaxy IR emission in the rest-frame MIR. 

\begin{figure}
    \centering
	\includegraphics[width=\columnwidth]{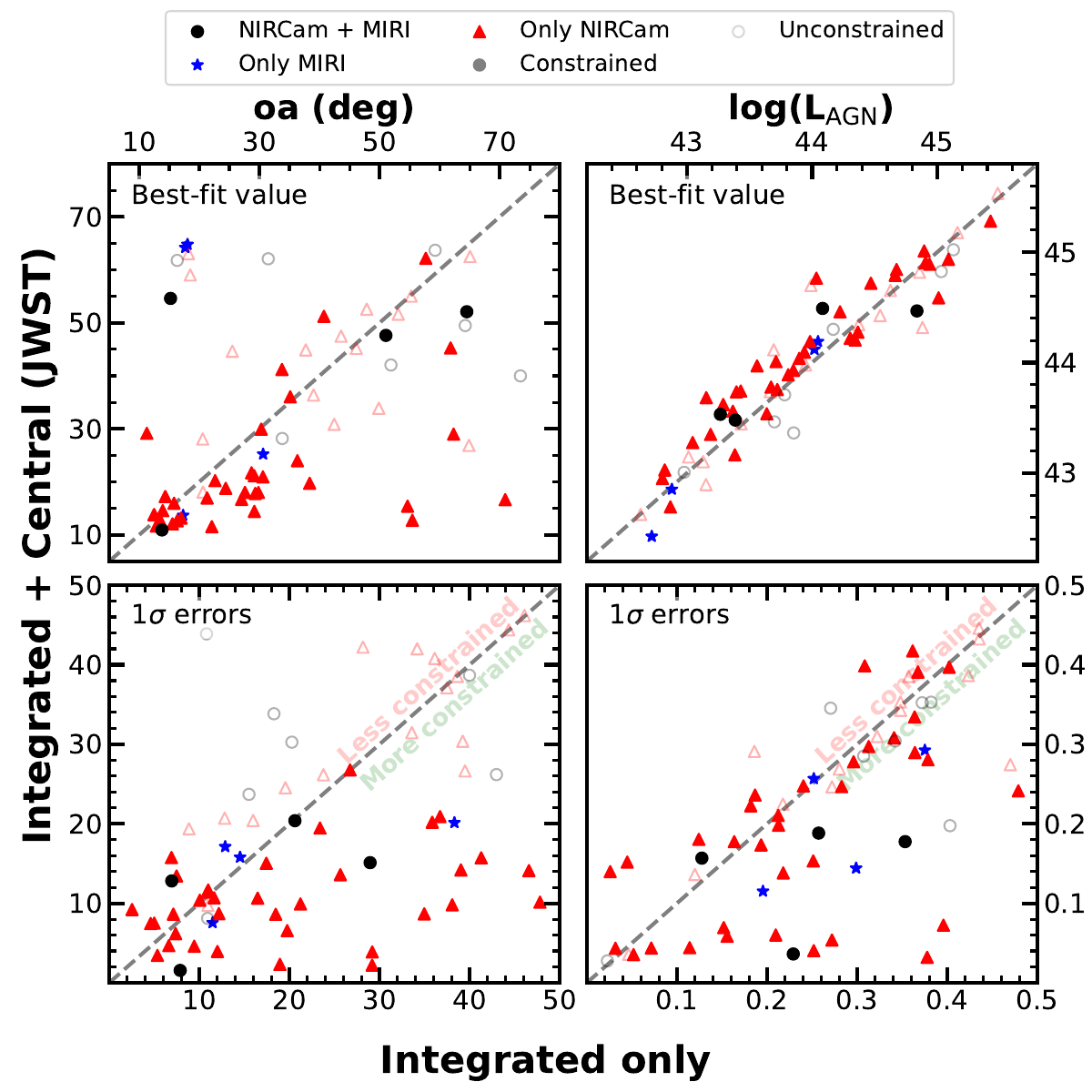}
    \caption{Fit results for two AGN parameters using only the integrated SED (x-axis), and using both integrated and central SEDs (y-axis) during SED fitting. The top row shows the best-fit values while the bottom row shows $1\sigma$ errors on those values. Different coloured symbols correspond to different \textit{JWST} coverage. The open symbols are sources where a particular parameter is unconstrained even with \textit{JWST} observations. The dashed line shows the line of perfect agreement between the two values. The terms ``more'' or ``less constrained'' in the figure refer to improvement or worsening of uncertainty on best-fit model after adding central JWST photometry.}
    \label{fig:jwst_improvement}
\end{figure}

We can demonstrate the overall benefits of the \textit{JWST} photometry by comparing, between the two sets of fits, the values of key parameters such as \Laccdisc\ and $oa$ (which is the basis for $CF$), including their uncertainties. Considering these two important AGN parameters, Figure \ref{fig:jwst_improvement} shows the best-fit parameter values and their equivalent $1\sigma$ errors for all objects in our sample, comparing integrated SED fits (x-axis) and fits with the central SED (y-axis). The best-fit values (top two panels) show some scatter between the two determinations, especially for $oa$, showing that the imposed priors, which are identical for both sets of fits, do not dominate our results. The higher scatter in $oa$, compared to that in \Laccdisc, suggests that our knowledge of the underlying nature of the AGN can change more significantly when central photometry are included, like in left column of Figure \ref{fig:example_fits2}, while these do not substantially alter the derived AGN power. We caution, however, that the AGN power is ultimately tied to \Lxray\ in both sets of fits. 
 
Looking at the uncertainties (lower two panels), we see that these are systematically reduced when using the \textit{JWST} central photometry during SED fitting, particularly for fits that are constraining (as defined in Section \ref{subsec:trendwithxray}) these parameters. Interestingly, we find no conclusive evidence that the degree of \textit{JWST} photometric coverage (denoted by different coloured symbols) plays a role in the constraining power of the fits. However, the coverage varies drastically even within the same subset (i.e., some ``only MIRI'' sources might have photometry from 2 filters while others might have 6 filters), hence it may be premature to make a definitive statement about the optimal level of \textit{JWST} coverage for this type of study.

\subsection{Caveats and future improvements}
\label{sub:improvements}
This work represents the most comprehensive \textit{JWST}-based multiwavelength study of a sizable sample of Seyfert-like AGN at cosmic noon, both in terms of the complexity of the models and the robustness of the SED fitting methods. Nevertheless, there are a few caveats that should be considered when attempting to place our results in the context of the general evolution of the AGN population. We present those caveats along with the mitigation steps we have taken wherever reasonable. 

AGN selection based on the $2-10$ keV X-ray luminosity (\Lxray) is known to be biased against the most obscured systems, even though this selection also gives us the largest and most diverse sample of AGN, across a wide range of obscuration, when compared to other selection methods \citep{Padovani2017}. Since the goal of this work is to demonstrate the power of \textit{JWST} in constraining AGN parameters in a wide variety of sources, we use X-ray selection to furnish ourselves with a representative set of systems for our analysis. Similarly-selected AGN in the local universe are available, though, as we discuss in Section \ref{subsec:cf_compare}, there are still some subtle differences in selection which may influence our conclusions about the lack of torus evolution.

The second caveat we note is our determination of the galaxy contribution in the central SED, as described in Section \ref{subsec:sed_modelling_method}. This galaxy contribution is fixed to a single number during the simultaneous SED fitting, which can be thought of as an extremely strong prior. Like any strong prior this can eliminate a large proportion of the possible parameter space during an SED fit, and drive it to a wrong result if the initial galaxy contribution is erroneously assigned. We have attempted to minimise this type of error by ascertaining the galaxy contribution at a rest wavelength of 1\mcron, at which the host galaxy light is most likely to dominate over the AGN. A possible alternative approach, which we have not explored in detail, is to rely on galaxy morphological decomposition to separate out the central point source emission from some parametrised model for the galaxy's light distribution. This method is highly reliant on the accuracy of the light profile model for the central galaxy emission (i.e., the galaxy bulge) and its variation with wavelength. Full morphological decomposition of our sample in all the \textit{JWST} images is beyond the scope of this current work, but in a future study, we will validate our approach against a robust set of structural fits for our sample.

Finally, while our final working set of 66 AGN is sizable for a sample of Seyfert-like AGN at cosmic noon, it still spans a broad range in redshift ($0.7 \lesssim z \lesssim 3.5$) and AGN luminosity. Given the diversity of AGN SEDs, even with a sample of 66 sources, it is statistically infeasible to further slice the sample on properties such as X-ray obscuration or luminosity, which may be required to investigate possible mechanisms driving the evolution (or lack thereof) of AGN tori. However, the focus of this work is to demonstrate the power of our novel SED fitting technique. We plan to expand our sample to a larger population in the future which would allow us to test various evolutionary scenarios in a statistically significant manner using the methods developed here. This includes the compilation of an expanded set of AGN in the larger \textit{JWST} fields now available (for e.g., COSMOS-Web \cite{Casey2023} which covers a much larger field with limited filters), and the use of alternative AGN selections, such as SED-based methods \citep[e.g.,][]{Andonie2022} which are less biased against obscured AGN.

\section{Conclusions}
\label{sec:conclusions}
In this work, we explore the properties of a set of AGN in the CEERS extragalactic field, building on the spectacular improvements that \textit{JWST} offers, particularly high spatial resolution imaging photometry in the NIR/MIR. 
We have developed a novel fitting method to simultaneously fit both the central and integrated galaxy-wide SEDs of AGN host galaxies using a robust Bayesian fitting framework. This approach allows the purer AGN emission from the central SED to be self-consistently revealed with better precision than in the past. We have demonstrated the improvements brought about by this approach by analysing the SEDs of a sample of 88 X-ray selected AGN between $0.7 \leq z \leq 3.5$.

\begin{itemize}
    \item The SED fitting method developed here leads to systematically tighter constraints on AGN parameters such as the torus covering factor $CF$ and \Laccdisc\ when the central SED from high-resolution \textit{JWST} imaging is folded into the analysis.
    \item Building on this, we constrained $CF$ in our sample, a first with \textit{JWST} data in a deep extragalactic field. We find that the high quality IR photometry of the centres of AGN hosts  also improves the overall SED fit, and is crucial in accurate and more precise modelling of the AGN itself. NIRCam photometry is critical for accurate estimates of the UV-optical AGN emission, while MIRI photometry is important for constraining the geometrical parameters of the torus. In many cases, the precise constraints on AGN shape afforded just by NIRCam observations can lead to significantly improved constraints on other AGN parameters.    
    \item From a working subsample of 66 sources with \Lxray\;$\geq 10^{43}$ \ergs, our results reveal a distribution of $CF$ that is clearly peaked at low values ($\approx 0.25$), and flattens beyond $CF=0.4$. At face value, this result is inconsistent with studies of the local ultrahard X-ray selected AGN population, for which the $CF$ distribution drops more rapidly at high $CF$ (a smaller fraction of systems with large tori). However, KS test reveals that these two distributions can not be considered inconsistent within generally accepted $< 5 \%$ confidence level.
    \item The qualitative differences we find against the local AGN population disappear when we use only the ``best-fit'' values of $CF$ from SED fits without accounting for the full posterior distribution. This means that the choices made during (such as shape of prior distribution) and after (such as method of interpreting posterior distribution) the SED fitting process can appear as misleading signs of ``evolution''.
    \item On the other hand, our distribution of $CF$ compares well to an intermediate redshift sample, also selected using similar criteria. The KS test was not valid for this due to modelling differences.
    \item For a sub-sample of 51 AGN for which there is a measure of \NH, due to X-ray spectral fits, we find that the distributions of covering fractions are identical for sources with log(\NH) $ \leq 22$ and log(\NH) $> 22$. This is at odds with the strong differences we find when we classify AGN into spectral types based on the relationship between torus inclination and opening angle. Future analysis which uses independent spectral information will help us investigate the trends with obscuration more robustly.
    \item Our fitting approach devised also yields distributions of AGN host galaxy properties which are in good agreement with previous work. This serves as valuable validation for our methodology.
\end{itemize}

High-resolution IR imaging from \textit{JWST} has opened up new possibilities for studying the fine details of the AGN population around cosmic noon, allowing us to explore, for the first time, the evolution of the torus in systems that are analogous to the Seyfert galaxies we see in the local Universe. We are only starting to extract the full value of these observations and our initial results are promising. The next steps will involve scaling these methods to the much larger number of AGN covered by current and future deep \textit{JWST} extragalactic surveys, towards a complete picture of the evolution of the nuclear regions of AGN over cosmic time.

\section*{Acknowledgements}

DHL would like to thank Chris Harrison for his invaluable discussions and comments throughout this work. DHL thanks Cristina Ramos Almeida for her constructive comments on the initial results of this work. Authors also thank Kohei Ichikawa for providing raw data files from I19, used for comparison in this work. Authors thank the anonymous referee and the editors for their constructive feedback. DHL acknowledges the UK Science and Technology Facilities Council (STFC) for support from grant No. ST/W006790/1. DJR acknowledges the support of the UK STFC through grant ST/X001105/1. This work makes use of observations taken by the CANDELS Multi-Cycle Treasury Program with the NASA/ESA \textit{HST}, which is operated by the Association of Universities for Research in Astronomy, Inc., under NASA contract NAS5-26555.

%%%%%%%%%%%%%%%%%%%%%%%%%%%%%%%%%%%%%%%%%%%%%%%%%%
\section*{Data Availability}
All the models and archival catalogues used in this work are available publicly at the links mentioned in the text. The data for all the tables and main figures in this article are available in electronic form in the supplementary material with this work. The aperture corrected photometric measurements from \textit{JWST} images (referred to as ``central SED'' in the text) along with a compilation of all the archival photometry (referred to as ``integrated SED'' in the text) are also available in supplementary material. Newcastle University repository with DOI:10.25405/data.ncl.28847105\footnote{https://doi.org/10.25405/data.ncl.28847105} contains full posterior distributions resulting from SED fitting, along with a notebook to read all different data product. The SED fitting code FortesFit used here is distributed as a python package via PyPI\footnote{https://pypi.org/project/fortesfit/2.0.0}. Any further data or assistance with the software developed here can be provided on request to the corresponding author.



\bibliographystyle{mnras}
\bibliography{example} % if your bibtex file is called example.bib

%%%%%%%%%%%%%%%%%%%%%%%%%%%%%%%%%%%%%%%%%%%%%%%%%%

%%%%%%%%%%%%%%%%% APPENDICES %%%%%%%%%%%%%%%%%%%%%

\appendix

\section{Sample properties}
\label{app:sample_prop}
We list some of the important properties of all 88 sources in our sample in Table \ref{tab:sample_prop}. These properties include counterpart coordinates, coordinates after recentring using IR images, the IR image used to do the recentring, and finally common properties in form of redshift, luminosity, and column density (whenever available). A full version of this table contains a number of useful columns that are not shown here due to space restrictions. These columns include a list of all the available filters for a given source, Rainbow ID of each of the source (as given in \citetalias{Nandra15_AEGISX}), collection of redshifts and luminosities from all sources mentioned in Section \ref{subsec:sample} along with the ``best'' redshifts and luminosities, and the source IDs of matched CANDELS and HELP counterparts. This can be found in electronic form in supplementary material provided with this paper.

\begin{table*}
	\centering
	\caption{Table of important properties of all 88 sources in our final sample. The columns are as follows: (1) XID of the source as given in \citetalias{Nandra15_AEGISX}, (2) \& (3) Right Ascension \& Declination of the AEGIS-X optical/IR counterpart, (4) \& (5) Right Ascension \& Declination of the source after re-centring (Section \ref{subsec:recentring}), (6) the IR image used for re-centring identified by its ``instrument\_filter\_pointing'', (7), (8) \& (9) best redshift (Section \ref{subsec:sample}) with its low and high values respectively, (10) best \Lxray\;estimate (Section \ref{subsec:sample}), (11) \& (12) \NH\;value \& its error as determined by X-ray spectral fitting in \citetalias{Buchner_AEGISX}. An electronic version of this table with additional columns is available as the supplementary material with this paper.}
	\label{tab:sample_prop}
    \begin{tabular}{|l|l|l|l|l|l|l|l|l|l|l|l|}
        \hline
        \hline
        XID & RA\_CP    & DE\_CP   & RA\_JW           & DE\_JW           & Recentring Obs. & $z$     & $z_{\rm lo}$ & $z_{\rm hi}$ & log(\Lxray)    & log(\NH)   & log(\NH) (err) \\
        (1) & (2) & (3) & (4) & (5) & (6) & (7) & (8) & (9) & (10) & (11) & (12) \\
        \hline
        \hline
        366 & 214.7321278 & 52.7519978 & 214.7321228 & 52.7519680 & nircam\_f444w\_4 & 1.417 & 1.269 & 1.546 & 43.26 &  &  \\
        368 & 214.7561808 & 52.7527047 & 214.7561397 & 52.7526637 & nircam\_f444w\_4 & 0.845 & 0.845 & 0.845 & 43.48 & 21.2 & 0.39 \\
        377 & 214.7517655 & 52.7618882 & 214.7517462 & 52.7618500 & nircam\_f444w\_4 & 2.634 & 2.47 & 2.798 & 43.81 & 23.2 & 0.13 \\
        378 & 214.7948301 & 52.762096 & 214.7948314 & 52.7620590 & nircam\_f444w\_10 & 0.956 & 0.956 & 0.956 & 42.96 & 23.1 & 1.92 \\
        381 & 214.7820047 & 52.7662107 & 214.7819447 & 52.7661219 & nircam\_f444w\_4 & 1.508 & 1.451 & 1.565 & 44.07 & 24.9 & 0.8 \\
        389 & 214.7198597 & 52.7719171 & 214.7198282 & 52.7718712 & miri\_f770w\_9 & 1.177 & 1.177 & 1.177 & 42.13 &  &  \\
        396 & 214.7885142 & 52.7759496 & 214.7885542 & 52.7758633 & nircam\_f444w\_4 & 1.285 & 1.065 & 1.504 & 43.16 & 21.6 & 1.65 \\
        400 & 214.7501368 & 52.7780655 & 214.7501042 & 52.7780704 & miri\_f560w\_9 & 1.508 & 1.266 & 1.75 & 43.81 & 24.5 & 0.64 \\
        409 & 214.7945957 & 52.7840803 & 214.7945754 & 52.7840136 & nircam\_f444w\_4 & 0.82 & 0.82 & 0.82 & 43.45 & 21.8 & 0.12 \\
        416 & 214.8184607 & 52.7863122 & 214.8183989 & 52.7863549 & nircam\_f444w\_4 & 2.042 & 1.947 & 2.137 & 44.8 & 25.3 & 0.57 \\
        421 & 214.8659286 & 52.7893678 & 214.8658518 & 52.7894564 & nircam\_f444w\_10 & 2.035 & 1.714 & 2.258 & 42.93 &  &  \\
        423 & 214.8477246 & 52.7909424 & 214.8477855 & 52.7909301 & nircam\_f444w\_10 & 0.738 & 0.738 & 0.738 & 43.64 & 23.8 & 0.11 \\
        424 & 214.9104163 & 52.7916564 & 214.9103852 & 52.7916308 & nircam\_f444w\_9 & 1.395 & 1.182 & 1.488 & 42.39 &  &  \\
        431 & 214.7945181 & 52.7975974 & 214.7945587 & 52.7976061 & nircam\_f444w\_4 & 1.911 & 1.911 & 1.911 & 44.28 &  &  \\
        432 & 214.8587324 & 52.7993053 & 214.8587763 & 52.7992950 & nircam\_f444w\_10 & 0.437 & 0.437 & 0.437 & 43.21 & 24.4 & 0.72 \\
        436 & 214.7881626 & 52.8020507 & 214.7881913 & 52.8020354 & nircam\_f444w\_4 & 1.503 & 1.343 & 1.584 & 43.26 &  &  \\
        443 & 214.8291774 & 52.8084356 & 214.8292434 & 52.8084363 & nircam\_f444w\_6 & 1.574 & 1.515 & 1.634 & 44.66 & 23.1 & 0.06 \\
        450 & 214.8634393 & 52.8155924 & 214.8634959 & 52.8156083 & nircam\_f444w\_6 & 1.683 & 1.497 & 1.87 & 43.5 & 21 & 0.74 \\
        455 & 214.8086218 & 52.8191692 & 214.8086520 & 52.8191855 & nircam\_f444w\_6 & 0.825 & 0.825 & 0.825 & 42.59 & 21.4 & 1.34 \\
        456 & 214.8770461 & 52.8194723 & 214.8771047 & 52.8195057 & nircam\_f444w\_9 & 1.204 & 1.119 & 1.256 & 42.93 &  &  \\
        457 & 214.8046948 & 52.8198353 & 214.8045661 & 52.8197736 & nircam\_f444w\_6 & 2.038 & 1.97 & 2.1 & 43.41 &  &  \\
        460 & 214.8122399 & 52.8248357 & 214.8123206 & 52.8250286 & nircam\_f444w\_6 & 1.197 & 1.197 & 1.197 & 42.29 &  &  \\
        463 & 214.7768027 & 52.8258763 & 214.7768442 & 52.8258792 & nircam\_f444w\_3 & 2.274 & 2.274 & 2.274 & 43.63 & 23.6 & 0.45 \\
        470 & 214.7494209 & 52.8295474 & 214.7494378 & 52.8294769 & nircam\_f444w\_3 & 1.769 & 1.552 & 1.986 & 43.12 & 21.1 & 1.14 \\
        471 & 214.7510839 & 52.8300523 & 214.7511244 & 52.8300821 & nircam\_f444w\_3 & 1.899 & 1.824 & 1.975 & 44.33 & 25 & 0.49 \\
        473 & 214.8441599 & 52.8314165 & 214.8442101 & 52.8314832 & miri\_f1000w\_8 & 1.199 & 1.199 & 1.199 & 43.16 & 24 & 0.74 \\
        474 & 214.7350258 & 52.8319353 & 214.7348543 & 52.8319023 & nircam\_f200w\_3 & 1.58 & 1.481 & 1.737 & 42.24 &  &  \\
        475 & 214.932894 & 52.8329538 & 214.9327697 & 52.8328909 & nircam\_f444w\_9 & 3.094 & 2.845 & 3.344 & 44.34 & 24.4 & 0.55 \\
        477 & 214.8707311 & 52.8331171 & 214.8707641 & 52.8331399 & nircam\_f444w\_6 & 2.317 & 2.317 & 2.317 & 44.32 & 21.1 & 0.64 \\
        480 & 214.7642277 & 52.8352886 & 214.7641975 & 52.8353344 & nircam\_f444w\_3 & 2.112 & 1.756 & 2.255 & 42.81 &  &  \\
        481 & 214.7857083 & 52.83608 & 214.7857094 & 52.8360724 & nircam\_f444w\_3 & 2.187 & 1.974 & 2.401 & 43.68 & 23.1 & 0.2 \\
        482 & 214.7552195 & 52.8367947 & 214.7552409 & 52.8367875 & nircam\_f444w\_3 & 3.465 & 3.465 & 3.465 & 44.13 & 21 & 0.63 \\
        485 & 214.885983 & 52.839642 & 214.8858278 & 52.8397405 & nircam\_f444w\_6 & 1.919 & 1.684 & 2.154 & 43.34 & 22.9 & 0.34 \\
        487 & 214.7853886 & 52.8402798 & 214.7853272 & 52.8402390 & nircam\_f444w\_3 & 1.314 & 1.239 & 1.389 & 44.35 & 25.3 & 0.58 \\
        488 & 214.9479055 & 52.8406401 & 214.9479649 & 52.8406266 & nircam\_f444w\_9 & 1.455 & 1.339 & 1.571 & 44.09 & 24 & 0.52 \\
        489 & 214.83716 & 52.8407187 & 214.8372974 & 52.8408553 & miri\_f1500w\_8 & 1.507 & 1.423 & 1.715 & 42.12 &  &  \\
        509 & 214.8987373 & 52.8525104 & 214.8987173 & 52.8524451 & nircam\_f444w\_6 & 2.138 & 1.911 & 2.364 & 43.53 & 22 & 1.3 \\
        511 & 214.8956059 & 52.8565158 & 214.8956193 & 52.8564934 & nircam\_f444w\_6 & 2.748 & 2.541 & 3.052 & 43.97 &  &  \\
        518 & 215.0126553 & 52.8587482 & 215.0125757 & 52.8587313 & nircam\_f356w\_8 & 1.933 & 1.775 & 2.091 & 43.76 & 21.8 & 0.67 \\
        523 & 214.9111389 & 52.8607545 & 214.9112262 & 52.8606802 & nircam\_f444w\_5 & 2.026 & 1.866 & 2.221 & 43.68 &  &  \\
        525 & 214.8538885 & 52.8614189 & 214.8539082 & 52.8613577 & nircam\_f444w\_3 & 1.634 & 1.462 & 1.807 & 43.09 & 23.1 & 0.34 \\
        530 & 214.9528146 & 52.8650349 & 214.9527420 & 52.8650030 & nircam\_f444w\_9 & 1.098 & 1.098 & 1.098 & 43.25 &  &  \\
        532 & 214.8505996 & 52.8664593 & 214.8505803 & 52.8664228 & nircam\_f444w\_3 & 2.235 & 1.985 & 2.485 & 44.02 & 24.8 & 1.44 \\
        538 & 214.7938011 & 52.8690413 & 214.7937458 & 52.8689766 & nircam\_f444w\_3 & 2.946 & 2.638 & 3.037 & 44.12 &  &  \\
        540 & 214.8153451 & 52.8709758 & 214.8153551 & 52.8709750 & nircam\_f444w\_3 & 0.452 & 0.452 & 0.452 & 43.38 & 25 & 0.55 \\
        544 & 214.9383186 & 52.8742464 & 214.9380774 & 52.8742962 & nircam\_f444w\_5 & 1.529 & 1.529 & 1.529 & 42.74 &  &  \\
        556 & 215.0059476 & 52.886126 & 215.0058745 & 52.8861043 & nircam\_f356w\_8 & 0.461 & 0.461 & 0.461 & 42.14 &  &  \\
        565 & 215.0292018 & 52.8954674 & 215.0291069 & 52.8954403 & nircam\_f444w\_8 & 1.097 & 1.097 & 1.097 & 43.9 & 23.6 & 0.07 \\
        575 & 215.0567661 & 52.9036141 & 215.0566865 & 52.9036008 & nircam\_f444w\_8 & 0.575 & 0.575 & 0.575 & 43.16 & 20.9 & 0.33 \\
        578 & 214.8502808 & 52.9055294 & 214.8502567 & 52.9055192 & nircam\_f444w\_2 & 1.032 & 1.032 & 1.032 & 43.07 &  &  \\
        582 & 214.8832197 & 52.9068988 & 214.8832143 & 52.9068239 & nircam\_f444w\_2 & 1.175 & 1.108 & 1.226 & 43.32 &  &  \\
        585 & 214.9316583 & 52.908708 & 214.9316179 & 52.9086928 & nircam\_f444w\_2 & 3.435 & 3.435 & 3.435 & 44.5 & 21.2 & 0.72 \\
        589 & 214.8592688 & 52.9115408 & 214.8592619 & 52.9115389 & nircam\_f444w\_2 & 2.252 & 2.111 & 2.393 & 43.85 & 21.4 & 0.91 \\
        599 & 214.9150199 & 52.9189083 & 214.9149405 & 52.9191838 & nircam\_f444w\_2 & 0.755 & 0.755 & 0.755 & 43.13 & 20.5 & 0.38 \\
        600 & 215.0416955 & 52.9199585 & 215.0418704 & 52.9199112 & nircam\_f444w\_8 & 0.702 & 0.651 & 0.733 & 42.51 &  &  \\
        \hline
        \hline
    \end{tabular}
\end{table*}

\begin{table*}
	\centering
	\caption{Table \ref{tab:sample_prop} continued.}
	\label{tab:sample_prop2}
    \begin{tabular}{|l|l|l|l|l|l|l|l|l|l|l|l|}
        \hline
        \hline
        XID & RA\_CP    & DE\_CP   & RA\_JW           & DE\_JW           & Recentring Obs. & $z$     & $z_{\rm lo}$ & $z_{\rm hi}$ & \Lxray    & \NH   & \NH(err) \\
        (1) & (2) & (3) & (4) & (5) & (6) & (7) & (8) & (9) & (10) & (11) & (12) \\
        \hline
        \hline
        612 & 215.079347 & 52.9342793 & 215.0792604 & 52.9342549 & nircam\_f444w\_7 & 2.607 & 2.52 & 2.695 & 44.16 &  &  \\
        616 & 214.9620111 & 52.9370994 & 214.9619478 & 52.9370880 & nircam\_f444w\_1 & 0.756 & 0.756 & 0.756 & 42.45 &  &  \\
        617 & 215.0303311 & 52.9375253 & 215.0302392 & 52.9374152 & nircam\_f356w\_8 & 2.478 & 2.333 & 2.609 & 44.18 &  &  \\
        618 & 214.9160839 & 52.937389 & 214.9160296 & 52.9373790 & nircam\_f444w\_2 & 0.745 & 0.745 & 0.745 & 43.26 & 22.9 & 0.06 \\
        623 & 215.0562126 & 52.9395732 & 215.0561275 & 52.9395412 & nircam\_f444w\_7 & 3.484 & 3.484 & 3.484 & 44.25 & 21.4 & 0.73 \\
        628 & 214.913172 & 52.9455714 & 214.9130926 & 52.9455147 & nircam\_f444w\_2 & 1.177 & 1.177 & 1.177 & 43.26 &  &  \\
        630 & 214.9420129 & 52.9463909 & 214.9421553 & 52.9464115 & nircam\_f444w\_1 & 2.218 & 2.061 & 2.362 & 44.15 &  &  \\
        634 & 215.0657925 & 52.9486951 & 215.0657020 & 52.9487024 & nircam\_f444w\_7 & 0.77 & 0.77 & 0.77 & 42.54 &  &  \\
        642 & 214.9554584 & 52.9513172 & 214.9553514 & 52.9512949 & nircam\_f444w\_1 & 1.282 & 1.282 & 1.282 & 43.19 &  &  \\
        650 & 215.0604226 & 52.9558321 & 215.0603624 & 52.9558352 & nircam\_f356w\_7 & 1.264 & 1.264 & 1.264 & 43.55 & 23.1 & 0.08 \\
        653 & 215.1216816 & 52.9574773 & 215.1215842 & 52.9574699 & nircam\_f444w\_7 & 1.352 & 1.308 & 1.404 & 42.91 &  &  \\
        656 & 215.0720536 & 52.9606367 & 215.0719765 & 52.9605857 & nircam\_f356w\_7 & 0.906 & 0.74 & 1.073 & 42.62 & 23.3 & 0.86 \\
        658 & 215.1214113 & 52.9613636 & 215.1213224 & 52.9613546 & nircam\_f444w\_7 & 1.405 & 1.405 & 1.405 & 43.92 & 20.6 & 0.35 \\
        660 & 214.9770304 & 52.9623194 & 214.9769552 & 52.9623247 & nircam\_f444w\_1 & 1.057 & 0.986 & 1.175 & 42.2 &  &  \\
        664 & 214.9308924 & 52.9634417 & 214.9308011 & 52.9634328 & nircam\_f444w\_1 & 1.104 & 1.104 & 1.104 & 43.5 & 23.4 & 0.09 \\
        683 & 214.947788 & 52.9749628 & 214.9478000 & 52.9749532 & nircam\_f444w\_1 & 2.018 & 1.9 & 2.136 & 44.49 & 24.4 & 0.33 \\
        687 & 215.116915 & 52.9781612 & 215.1168094 & 52.9781596 & nircam\_f444w\_7 & 0.871 & 0.871 & 0.871 & 43.66 & 21.6 & 2.05 \\
        689 & 215.0863917 & 52.9783961 & 215.0863192 & 52.9783888 & nircam\_f356w\_7 & 0.631 & 0.631 & 0.631 & 42.67 & 22.4 & 1.97 \\
        693 & 215.1552642 & 52.9808399 & 215.1551644 & 52.9808412 & nircam\_f444w\_7 & 2.272 & 2.272 & 2.272 & 43.79 & 20.9 & 0.49 \\
        694 & 215.1690174 & 52.9807773 & 215.1689765 & 52.9807546 & nircam\_f356w\_7 & 2.276 & 2.217 & 2.334 & 44.28 & 24.1 & 1.37 \\
        699 & 215.123927 & 52.9876707 & 215.1240274 & 52.9877304 & nircam\_f444w\_7 & 0.892 & 0.75 & 1.035 & 43.7 & 24.5 & 1.12 \\
        704 & 215.1400326 & 52.988158 & 215.1398775 & 52.9881159 & nircam\_f444w\_7 & 1.297 & 1.297 & 1.297 & 43.92 & 25 & 0.85 \\
        710 & 215.1094003 & 52.9918613 & 215.1093177 & 52.9918565 & nircam\_f356w\_7 & 1.525 & 1.49 & 1.643 & 42.37 &  &  \\
        712 & 215.0110796 & 52.9925549 & 215.0110740 & 52.9925576 & nircam\_f444w\_1 & 1.619 & 1.46 & 1.779 & 44.15 & 25 & 0.73 \\
        715 & 215.0118685 & 52.9955537 & 215.0117557 & 52.9955146 & nircam\_f444w\_1 & 0.915 & 0.915 & 0.915 & 43.93 & 25.1 & 0.61 \\
        717 & 215.1156451 & 52.9968831 & 215.1155786 & 52.9968692 & nircam\_f356w\_7 & 1.008 & 0.899 & 1.116 & 43.12 & 20.6 & 0.3 \\
        721 & 215.0086013 & 52.9973543 & 215.0085241 & 52.9973333 & nircam\_f444w\_1 & 0.912 & 0.912 & 0.912 & 44.02 & 25.1 & 0.73 \\
        723 & 215.053268 & 52.9980955 & 215.0533054 & 52.9980976 & miri\_f770w\_2 & 1.356 & 1.185 & 1.571 & 42.18 &  &  \\
        726 & 215.0455644 & 52.9995467 & 215.0455022 & 52.9995459 & nircam\_f444w\_1 & 1.716 & 1.558 & 1.874 & 43.65 & 21.1 & 1.02 \\
        742 & 215.0233929 & 53.010207 & 215.0233290 & 53.0101975 & nircam\_f444w\_1 & 1.644 & 1.644 & 1.644 & 44.45 & 21.1 & 0.53 \\
        783 & 215.1630247 & 53.031554 & 215.1626466 & 53.0315079 & miri\_f770w\_1 & 1.129 & 1.072 & 1.339 & 42.71 &  &  \\
        832 & 215.1420123 & 53.0575824 & 215.1418923 & 53.0575428 & miri\_f770w\_1 & 1.459 & 1.396 & 1.522 & 43.94 & 23.3 & 0.08 \\
        844 & 215.1548887 & 53.0625976 & 215.1548148 & 53.0625938 & miri\_f770w\_1 & 0.718 & 0.718 & 0.718 & 43.15 & 24.4 & 1.3 \\
        \hline
        \hline
    \end{tabular}
\end{table*}

\section{Distinguishing constrained and unconstrained parameters}
\label{app:klderror}
Within the Bayesian fitting framework here, we consider a parameter to be ``constrained'' if its posterior distribution differs significantly from the prior, i.e. the fit should yield a measurable gain in information. Conversely, if the posterior distribution closely resembles the prior then the parameter is said to be ``unconstrained''. Note that the ``updating of the knowledge'' can be in form of shrinkage in prior distribution (i.e. improved uncertainty) or shifting of the median parameter value compared to the prior distribution. We adopt these definitions of constrained/unconstrained throughout this work and use Kullback-Leibler Divergence (KLD) to quantify information gained by an SED fit. KLD is a statistical metric that quantifies the dissimilarity between two distributions.

\begin{figure*}
    \centering
	\includegraphics[width=0.95\textwidth]{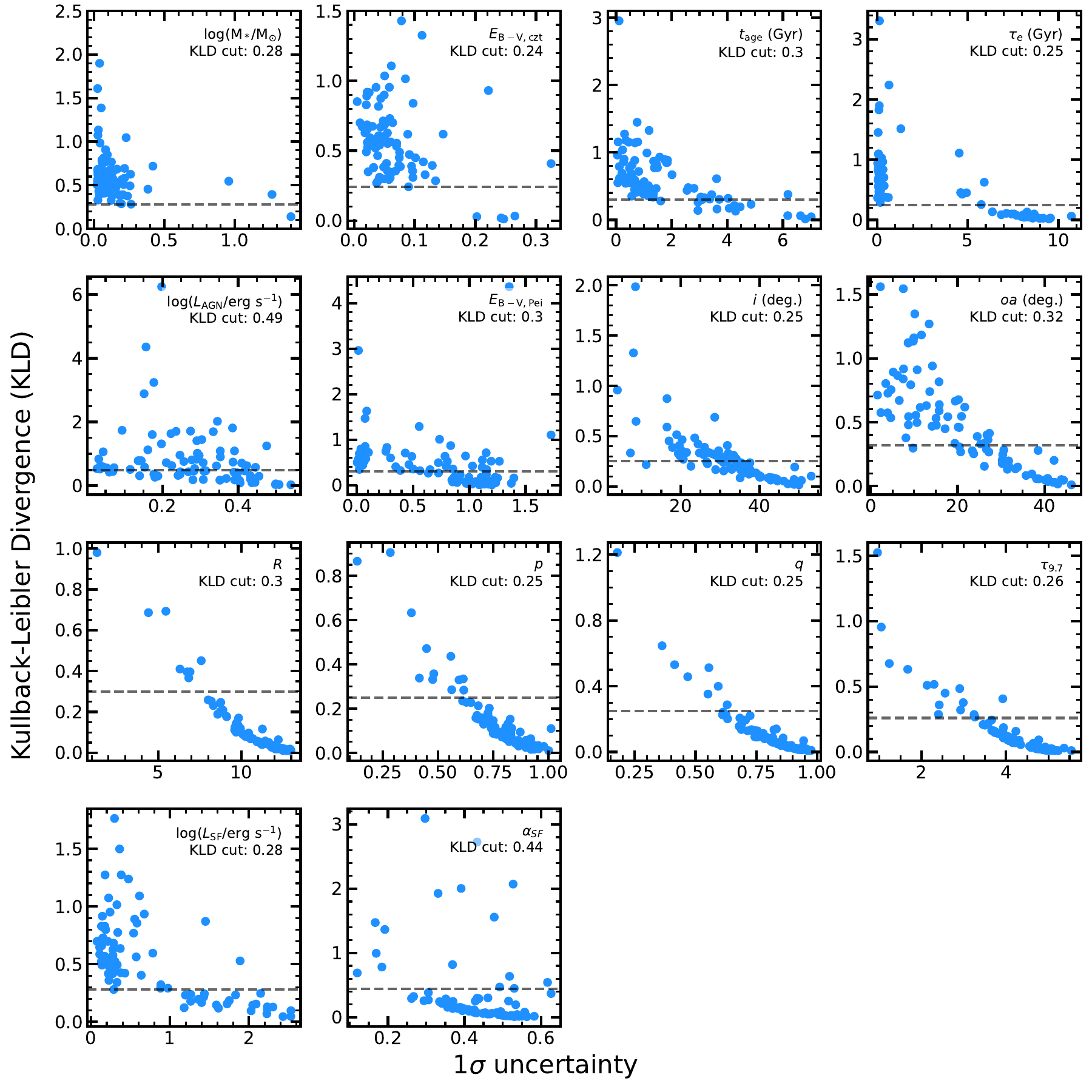}
    \caption{Values of Kullback-Leibler Divergence (KLD) plotted against the absolute 1$\sigma$ uncertainty on all the free model parameters. The black dashed line shows the KLD cutoff we have used to separate ``constrained'' parameters from ``unconstrained'' parameters for a given source, this threshold value is also mentioned in top right corner of each panel}.
    \label{figapp:kld_error}
\end{figure*}

Determining the threshold KLD value to distinguish constrained/unconstrained parameters is not trivial and the threshold differs from parameter to parameter because the numerical value of KLD depends on the underlying distributions being tested. We tackle this by examining the plot of KLD against the absolute uncertainty on each parameter shown in Figure \ref{figapp:kld_error}. Ideally, we expect a negative correlation between these two; the KLD increases as the uncertainty on the parameter decreases (i.e. posterior becomes narrower) for a fixed prior. In practice, we often observe an ``elbow'' in this plot which divides the prior dominated posteriors (unconstrained) where we see minimal change in KLD over a large range of uncertainty, and the likelihood dominated posteriors (constrained) where we see KLD increase rapidly over a small range of uncertainty. When we observe this elbow, we set the KLD threshold such that it divides these two regions of the plot. An example of this is log(\Lsf). However, for parameters where these regions are not visually separable, we select the threshold KLD such that there is significant scatter in uncertainty for KLD values above the threshold. Examples of this are log(\Laccdisc), $R$, $p$, $q$, etc. This approach takes into account varying priors for parameters like \Laccdisc. It also ensures that the two distributions are different enough no matter the size and shape of the prior distribution.

A more intuitive metric such as setting a threshold on relative parameter error ($\sigma_{\rm par}/\rm{par}$), or measuring the shrinkage of credible interval between prior and posteriors are not appropriate for distinguishing constrained/unconstrained parameters beyond those with uniform priors because they only measure the shrinkage of the distribution and not the shift. We demonstrate this with an example. Consider a parameter which has a Gaussian prior with $\mu = 4$ and $\sigma=2$. A particular SED fitting task has returned a posterior which is nearly identical to the prior ($\mu\approx4,\;\sigma\approx2$). The parameter should be considered ``unconstrained'' in this case because there is minimal information gained from SED fitting. Another SED fitting task has returned a posterior with a significantly different mean but a larger standard deviation (say, $\mu\approx1,\;\sigma\approx4$). The information gained in this case is more and hence the parameter should be considered ``constrained''. However, measures such as relative error and credible interval shrinkage would result in the first case being classified as ``constrained'' whereas the second would be classified as ``unconstrained''. Therefore, a statistical measure of dissimilarity of two distributions is required to identify constrained/unconstrained parameters as we have defined them in this work.

We caution the reader against an obvious pitfall of using this method. Namely, it fails in the cases when the supplied prior distribution is exactly the ``true'' value of that parameter. The posterior distribution in that case should ideally be same as the prior distribution and there should be no information gain. However, this is a rare occurrence in our context, and is a worthy trade-off as long as we do not over-interpret the posteriors of parameters with non-uniform prior distributions. We therefore highlight the non-uniform prior distributions when discussing these parameter (e.g. \Laccdisc) in the main text. Moreover, our approach is more conservative in the sense that it may classify a few of constrained parameters as unconstrained, but it avoids falsely classifying an unconstrained parameters as constrained in most cases which is a downside of alternative methods.

\section{Best-fit parameters}
\label{app:bestfit}
Table \ref{apptab:bestfit} gives the ``best-fit'' values, defined by the median of posterior distribution, of each of the free parameters in our fits when using both integrated and the central \textit{JWST} SEDs (i.e. result of the second step of the two step SED fitting procedure described in \ref{subsub:finalSEDmodel}). While these values can be useful for quick lookup, we advise readers to interpret these in light of the $1\sigma$ errors provided in the electronic version of the table at the least or ideally in light of the full posterior distributions provided in Newcastle University repository with DOI:10.25405/data.ncl.28847105. The electronic version of the table also contains the KLD values associated with each parameter along with the threshold KLD values used here to distinguish constrained parameters from unconstrained ones.

\begin{landscape}
\begin{table}
	\centering
	\caption{The table of best-fit values (defined as the median of the appropriate posterior distribution) for 42 out of 88 sources in our final sample. An extended table with all 88 sources with the $1\sigma$ errors and KLD values is available in electronic form as supplementary material to this paper. Column (1) is the source ID used throughout this work while rest of the columns are the free parameters in our model as defined in, and in the same unit as, Table \ref{tab:model_params}.}
	\label{apptab:bestfit}
    \begin{tabular}{|l|l|l|l|l|l|l|l|l|l|l|l|l|l|l|}
        \hline
        \hline
        XID & \logMstar & $E_{\rm B-V, czt}$ & \Agyr & \Tgyr & log($L_{\rm AGN}$) & $E_{\rm B-V, Pei}$ & $i$ & $oa$ & $R$ & $p$ & $q$ & \opticaldepth & log($L_{\rm SF}$) & $\alpha_{\rm SF}$ \\
        (1) & (2) & (3) & (4) & (5) & (6) & (7) & (8) & (9) & (10) & (11) & (12) & (13) & (14) & (15) \\
        \hline
        \hline
        366 & 10.7108 & 0.3562 & 0.5159 & 0.1468 & 43.0064 & 0.8979 & 57.7627 & 42.0476 & 19.7356 & 0.7536 & 0.755 & 7.4993 & 44.3872 & 1.9631 \\
        368 & 10.407 & 0.0288 & 1.7887 & 0.1763 & 44.4904 & 0.2796 & 13.8175 & 10.9261 & 17.6196 & 0.4623 & 0.842 & 6.7695 & 44.871 & 1.8233 \\
        377 & 11.1852 & 0.18 & 8.0044 & 6.2945 & 44.3013 & 0.4548 & 26.1724 & 28.1821 & 19.9088 & 0.7034 & 0.7195 & 6.9236 & 45.4924 & 1.6698 \\
        378 & 10.8588 & 0.2637 & 3.1318 & 6.202 & 42.7343 & 0.689 & 33.0436 & 20.306 & 19.7635 & 0.6505 & 0.7529 & 7.0561 & 44.6363 & 1.9215 \\
        381 & 10.8212 & 0.4592 & 6.7883 & 8.6443 & 44.7183 & 0.5871 & 23.0689 & 12.391 & 17.4386 & 1.0439 & 0.7575 & 7.7611 & 45.8293 & 1.9289 \\
        389 & 10.8657 & 0.4339 & 0.9245 & 6.4184 & 41.7331 & 0.7093 & 39.9521 & 45.6448 & 20.4695 & 0.7362 & 0.7387 & 6.8945 & 45.6111 & 1.7822 \\
        396 & 10.742 & 0.3865 & 0.4967 & 0.1325 & 42.8956 & 1.2364 & 28.3673 & 26.8782 & 21.183 & 0.7661 & 0.906 & 6.5587 & 44.6748 & 2.1496 \\
        400 & 10.4942 & 0.0415 & 6.0175 & 0.364 & 44.1904 & 0.6163 & 22.4701 & 13.6861 & 18.9346 & 0.6228 & 0.7238 & 6.7048 & 44.663 & 1.9866 \\
        409 & 11.176 & 0.0985 & 6.2555 & 1.0847 & 43.7351 & 1.1227 & 25.9593 & 18.011 & 19.6802 & 0.7789 & 0.7525 & 6.8218 & 44.5573 & 1.9558 \\
        416 & 10.8614 & 0.3626 & 9.7221 & 4.5844 & 44.5866 & 3.2504 & 28.7005 & 14.4257 & 20.5469 & 0.552 & 0.6073 & 7.0929 & 45.5766 & 1.7353 \\
        421 & 10.3979 & 0.2379 & 7.7222 & 1.0398 & 44.0516 & 0.9777 & 19.0778 & 41.1059 & 20.8371 & 0.9046 & 0.7512 & 7.2198 & 44.4264 & 2.0115 \\
        423 & 11.121 & 0.1276 & 3.4176 & 0.3497 & 43.9298 & 1.2038 & 27.3986 & 21.7169 & 19.2256 & 0.6496 & 0.7727 & 7.1331 & 45.2365 & 1.9851 \\
        424 & 10.7671 & 0.0143 & 8.6306 & 1.4494 & 42.0835 & 0.5537 & 38.6514 & 37.1122 & 19.7667 & 0.6988 & 0.6162 & 7.3261 & 43.054 & 1.9894 \\
        431 & 10.4311 & 0.1798 & 3.5555 & 6.4099 & 44.2201 & 0.2947 & 23.3974 & 13.8182 & 19.4887 & 0.482 & 0.8143 & 6.5167 & 43.3225 & 1.9848 \\
        432 & 11.0356 & 0.0633 & 4.405 & 0.1176 & 42.6931 & 1.2116 & 29.979 & 15.9697 & 19.7042 & 0.6283 & 0.7797 & 6.9134 & 42.6421 & 1.9873 \\
        436 & 10.858 & 0.3022 & 1.0663 & 0.404 & 43.1457 & 0.7071 & 60.455 & 44.6349 & 21.7971 & 0.6879 & 0.5508 & 8.1801 & 44.8802 & 1.9817 \\
        443 & 11.8072 & 0.1696 & 8.4405 & 1.2285 & 44.4678 & 0.9539 & 65.0837 & 54.6058 & 19.3096 & 1.2229 & 0.9296 & 6.4212 & 44.5105 & 1.9428 \\
        450 & 10.76 & 0.2479 & 7.6644 & 10.1923 & 43.3634 & 0.7033 & 56.7682 & 39.9897 & 21.0526 & 0.5961 & 0.4605 & 8.5787 & 45.017 & 1.9937 \\
        455 & 11.3889 & 0.0817 & 6.6854 & 1.2932 & 42.5624 & 0.658 & 32.4115 & 44.979 & 19.3836 & 0.7762 & 0.754 & 7.4128 & 44.7428 & 2.1596 \\
        456 & 11.1529 & 0.339 & 5.7527 & 3.8928 & 42.8002 & 0.7181 & 32.2323 & 23.6557 & 19.7123 & 0.6359 & 0.7797 & 6.8523 & 45.2235 & 1.4537 \\
        457 & 10.8673 & 0.1492 & 8.9173 & 9.7653 & 44.1155 & 1.5562 & 26.6694 & 30.8582 & 20.1465 & 0.7914 & 0.7564 & 6.9036 & 46.0237 & 1.2785 \\
        460 & 11.0906 & 0.793 & 2.5152 & 8.1736 & 43.3687 & 0.0363 & 13.6035 & 44.8007 & 21.1895 & 0.6416 & 0.9566 & 6.0728 & 45.837 & 2.0914 \\
        463 & 10.673 & 0.0436 & 1.0891 & 0.3615 & 43.4637 & 1.3921 & 56.4243 & 62.0777 & 22.866 & 0.7694 & 0.5981 & 7.847 & 43.4322 & 1.9932 \\
        470 & 10.5688 & 0.0147 & 1.9697 & 0.2833 & 43.4796 & 0.7782 & 21.7917 & 52.0953 & 21.7245 & 0.6883 & 0.7699 & 6.6907 & 44.1853 & 1.989 \\
        471 & 11.8419 & 0.4529 & 8.3092 & 10.8172 & 44.8243 & 0.7538 & 55.8973 & 61.7716 & 20.4785 & 0.6213 & 0.3066 & 9.5101 & 46.0048 & 2.0052 \\
        473 & 11.1561 & 0.161 & 7.493 & 1.6945 & 42.8523 & 0.8722 & 47.9491 & 25.2265 & 18.8728 & 0.7001 & 0.8935 & 7.4935 & 45.1738 & 2.015 \\
        474 & 11.478 & 0.1199 & 8.474 & 1.6096 & 42.0042 & 1.4571 & 39.508 & 54.9991 & 22.4721 & 0.4836 & 0.7392 & 5.6842 & 45.5296 & 2.0997 \\
        475 & 11.3516 & 0.2128 & 3.0317 & 8.7104 & 44.8179 & 0.906 & 55.9851 & 47.4788 & 20.8831 & 0.6828 & 0.6445 & 7.3779 & 44.1848 & 1.9811 \\
        477 & 9.1376 & 0.1156 & 5.5113 & 8.0036 & 44.8913 & 0.027 & 26.4708 & 17.2423 & 19.0653 & 0.5683 & 0.7863 & 6.6508 & 45.1575 & 1.9853 \\
        480 & 11.0762 & 0.0079 & 8.8822 & 1.3945 & 42.5846 & 0.7257 & 36.4022 & 31.4685 & 19.3947 & 0.6827 & 0.6496 & 6.2927 & 43.2587 & 1.9955 \\
        481 & 10.5641 & 0.3483 & 0.5287 & 0.3098 & 43.7366 & 1.539 & 57.1777 & 63.0512 & 17.0426 & 0.6937 & 0.8905 & 8.1758 & 43.4199 & 1.7842 \\
        482 & 9.5875 & 0.0141 & 0.1432 & 6.9684 & 45.0217 & 0.0037 & 6.3003 & 63.6907 & 26.9626 & 1.0657 & 0.9438 & 7.9442 & 45.7024 & 2.0125 \\
        485 & 10.6966 & 0.3138 & 1.7814 & 8.2126 & 43.4453 & 0.8152 & 48.8152 & 44.8881 & 20.2452 & 0.7148 & 0.6915 & 7.333 & 45.8078 & 1.3471 \\
        487 & 10.6087 & 0.0366 & 3.4058 & 0.3796 & 43.8898 & 1.0568 & 76.2632 & 29.1564 & 26.3172 & 1.0179 & 0.2462 & 10.5359 & 43.0591 & 1.9722 \\
        488 & 11.1666 & 0.2816 & 1.027 & 0.1436 & 44.3174 & 1.5688 & 13.3831 & 59.0322 & 22.109 & 0.4599 & 0.9601 & 6.0742 & 45.9452 & 1.7199 \\
        489 & 10.1315 & 0.2854 & 0.4343 & 8.5486 & 41.7933 & 0.6769 & 34.3938 & 49.9273 & 19.9335 & 0.7827 & 0.8153 & 6.9803 & 44.4543 & 1.8949 \\
        509 & 11.4072 & 0.2168 & 10.1823 & 4.7818 & 44.6994 & 2.1531 & 23.1417 & 33.8848 & 19.8112 & 0.7972 & 0.7895 & 7.0648 & 46.1377 & 2.0042 \\
        511 & 10.4478 & 0.1808 & 5.1291 & 8.2632 & 45.011 & 0.3646 & 20.6711 & 11.6678 & 19.1064 & 0.3026 & 0.7851 & 6.0212 & 43.9512 & 1.991 \\
        518 & 10.5418 & 0.2272 & 3.4683 & 8.5049 & 43.9772 & 0.7686 & 59.7035 & 52.5734 & 21.6853 & 0.7493 & 0.5247 & 8.5541 & 43.6706 & 2.0102 \\
        523 & 10.8842 & 0.265 & 2.4485 & 7.5462 & 44.0377 & 1.3719 & 24.5584 & 15.4143 & 18.8488 & 0.6366 & 0.7771 & 6.792 & 45.246 & 1.9481 \\
        525 & 11.1195 & 0.8002 & 8.069 & 0.4666 & 43.5366 & 0.9007 & 64.2327 & 62.1651 & 20.0898 & 0.8889 & 0.8692 & 5.1829 & 43.9754 & 1.9652 \\
        530 & 10.8268 & 0.0344 & 2.6695 & 0.3404 & 43.3493 & 0.6863 & 27.7151 & 17.9963 & 20.1806 & 0.6395 & 0.7758 & 6.7186 & 44.461 & 2.0113 \\
        \hline
        \hline
    \end{tabular}
\end{table}
\end{landscape}

\section{Point estimates vs. posterior distributions from SED fits}
\label{app:pointest}
An important consideration before interpreting differences in population distribution as the signs of evolution in $CF$, and by extension a torus evolution in the underlying populations of AGN, is the effect of choices made during the SED fitting and in reporting its results. Specifically, we note that the SED fitting method used in \citetalias{Ichikawa19} does not give a full posterior distribution of each model parameter, but only a single best-fit value for $CF$, extrapolated from the SKIRTOR-based fitting function. In essence, these are ``point estimates'' of $CF$, equivalent to a best-fit value. In a similar vein, the distribution constructed from the individual best-fit values (defined by the median of the posterior distribution) of $CF$ for our AGN can also be interpreted as an equivalent distribution of ``point estimate''. 

\begin{figure}
    \centering
	\includegraphics[width=\columnwidth]{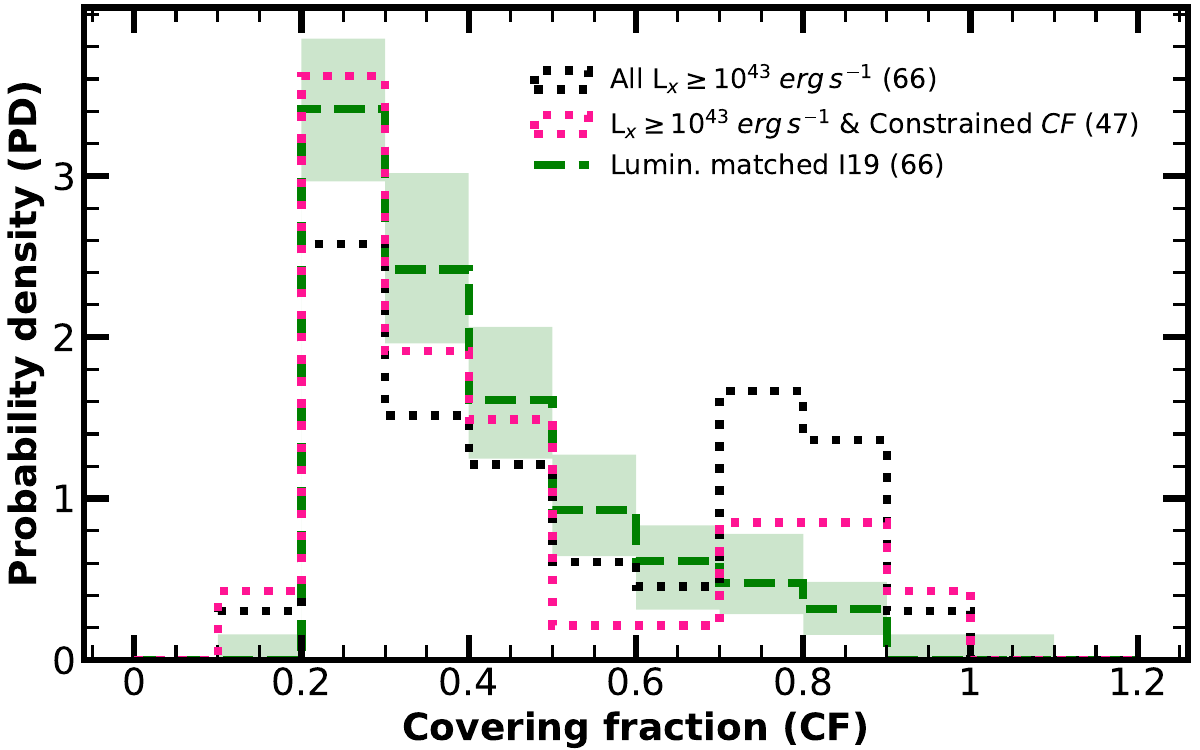}
    \caption{Covering fraction ($CF$) distributions obtained using ``point estimates'' i.e. a single best-fit value, defined as the median of posterior distribution, from all our fits compared with luminosity matched \protect\cite{Ichikawa19} sample. The black dotted line is from all sources with \Lxray\;$\geq 10^{43}$ \ergs, and the pink dotted line is from further filtering of these sources to only include those with constrained $CF$ as defined in Section \ref{subsec:trendwithxray}. The green dashed line is luminosity matched sample from \protect\citetalias{Ichikawa19} (the matching was done for all 66 sources with \Lxray\;$\geq 10^{43}$ \ergs), identical to that in Figure \ref{fig:CFcompare_others}}
    \label{fig:point_vs_post}
\end{figure} 

The Figure \ref{fig:point_vs_post} presents $CF$ distributions obtained using point estimates for two slightly different populations for our sample, to allow a fairer comparison to that the \citetalias{Ichikawa19} $CF$ distribution. The black dotted line shows the distribution from point estimates for all 66 sources with \Lxray\;$\geq 10^{43}$ \ergs. Recall that this subset also includes a number of AGN for which we have determined that there is no robust constraint on $CF$, and for which the point estimate is close to the median value set by the uninformative prior on the opening angle ($45^{\circ}$, or $CF \approx 0.7$). Not surprisingly, this is where we also see a small peak in this distribution. One may argue that objects without constraints should not be included in a comparison between distributions based on point-estimates.

Therefore, we also show the pink dotted line, the distribution for only those AGN from our sample for which $CF$ is constrained, as defined in Section \ref{subsec:trendwithxray}, and shown by filled circles in $oa$ panel of Figure \ref{fig:population_stats}. Many of the sources in the high $CF$ secondary peak of the black dotted histogram fall out of the sample when the unconstrained sources are excluded. 

The constrained $CF$ distribution is much more consistent with \citetalias{Ichikawa19} distribution (compare the pink dotted line and the green dashed line). Given this similarity, we conclude that the apparent sign of evolution in the amount of obscuration by AGN torus could be the result of systematics rather than any underlying astrophysical mechanisms. If so, any evolution in the $CF$ distributions is probably minor, and we would require a significantly larger sample of AGN selected in similar ways to make any statistical conclusions regarding the cosmic evolution of $CF$.

Similar to the main text, we performed a KS test for different pairs of distribution. For the pair of all sources (black dotted line) and \citetalias{Ichikawa19}, the statistic and p-values were 0.2273 and 0.066 respectively. For the pair of only constrained (with \Lxray\;$\geq 10^{43}$ \ergs) and \citetalias{Ichikawa19}, the statistic and p-values were 0.1702 and 0.5803. The p-values > 0.05 suggest that while neither of these can be considered to be statistically from different underlying distributions, the constrained distribution is significantly more similar to \citetalias{Ichikawa19} as suspected from visual examination.

The largest issue with relying on distributions based only on point estimates is that these estimates can significantly underestimate the systematic uncertainties in parameters, especially, in the case of $CF$, in sources with weak AGN contributions and limited MIR photometry. Point estimates can also be misleading in cases of a highly degenerate parameter spaces when using models that do not capture the full variation in SED shapes found in the data. The Bayesian inference used in this work mitigates some of these issues. With this in mind, our default distribution of $CF$ includes the sources determined to have unconstrained $CF$, in addition to the sources with constrained $CF$, using their full posterior distributions to get the population level statistics (i.e., the solid black line and shaded regions in Figure \ref{fig:CFcompare_others}). 

%%%%%%%%%%%%%%%%%%%%%%%%%%%%%%%%%%%%%%%%%%%%%%%%%%

% Don't change these lines
\bsp	% typesetting comment
\label{lastpage}
\end{document}